\documentclass[usenatbib]{mn2e}
\usepackage{color,hyperref}
\definecolor{darkblue}{rgb}{0.3,0.3,1.0}
\hypersetup{colorlinks,breaklinks,
            linkcolor=darkblue,urlcolor=darkblue,
            anchorcolor=darkblue,citecolor=darkblue}
\usepackage{amsmath,amssymb}

\DeclareSymbolFont{cmletters}{OML}{cmm}{m}{it}
\DeclareMathSymbol{v}{\mathalpha}{cmletters}{"76}

\usepackage{graphicx}

\defcitealias{mtb12b}{MTB12a}

\newcommand{\be}{\begin{equation}}
\newcommand{\ee}{\end{equation}}
\newcommand{\bea}{\begin{eqnarray}}
\newcommand{\eea}{\end{eqnarray}}

\title[Star Formation in the Galactic centre]
{Star Formation in the vicinity of Nuclear Black Holes: 
Young Stellar Objects close to Sgr~A*}
\author[Jalali,~B. et al.]
{Jalali,~B.$^{1}$\thanks{E-mail:\texttt{bjalali@ph1.uni-koeln.de}},  
Pelupessy, F. I.$^{2}$, Eckart, A.$^{1,3}$, Portegies Zwart, S.$^{2}$, Sabha, N.$^{1,3}$,
\newauthor Borkar, A.$^{1,3}$, Moultaka, J.$^{4}$, Mu\v{z}i\'{c}, K.$^{5}$, Moser, L.$^{1}$
\vspace{0.4cm}\\
$^{1}$ I. Physikalisches Institut, Universit\"at zu K\"oln, Z\"ulpicher Stra$\beta$e 77, D-50937 K\"oln, Germany\\
$^{2}$ Leiden Observatory, Leiden University, PO Box 9513, 2300 RA, Leiden, The Netherlands\\
$^{3}$ Max-Planck-Institut f\"ur Radioastronomie, Auf dem H\"ugel 69, 53121 Bonn, Germany\\
$^{4}$ IRAP, Observatoire Midi-Pyr\'{e}n\'{e}es, CNRS, Universit\'{e} Toulouse III, 14 avenue Edouard Belin, 31000 Toulouse, France\\
$^{5}$ European Southern Observatory, Alonso de C\'{o}rdova 3107, Casilla 19, Santiago, Chile
}

\begin{document}

\date{Accepted 2014 July 23. Received 2014 July 23; in original form 2013 October 01}
\pagerange{\pageref{firstpage}--\pageref{lastpage}} \pubyear{2013}
\maketitle

\label{firstpage}

\begin{abstract}

It is often assumed that the strong gravitational field of a super-massive black hole 
disrupts an adjacent molecular cloud preventing classical star formation in the deep 
potential well of the black hole. Yet, young stars have been observed across the entire 
nuclear star cluster of the Milky Way including the region close ($<$0.5~pc) to the 
central black hole, Sgr~A*.

Here, we focus particularly on small groups of young stars, such as IRS~13N located 
0.1~pc away from Sgr~A*, which is suggested to contain about five embedded massive young 
stellar objects ($<$1~Myr).
We perform three dimensional hydrodynamical simulations to follow the evolution of 
molecular clumps orbiting about a $4\times10^6~M_{\odot}$ black hole, to constrain the 
formation and the physical conditions of such groups.

The molecular clumps in our models assumed to be isothermal containing 100 $M_{\odot}$
in $<$0.2~pc radius. Such molecular clumps exist in the circumnuclear disk of the Galaxy.

In our highly eccentrically orbiting clump, the strong orbital compression of the clump 
along the orbital radius vector and perpendicular to the orbital plane causes 
the gas densities to increase to values higher than the tidal density of Sgr~A*, which 
are required for star formation.

Additionally, we speculate that the infrared excess source G2/DSO approaching Sgr~A* on
a highly eccentric orbit could be associated with a dust enshrouded star that may have 
been formed recently through the mechanism supported by our models.

\end{abstract}

\begin{keywords}  
Star Formation: IRS~13N association, ISM - Galactic centre, Black Holes: Sgr~A* - Methods: Numerical, SPH simulations, AMUSE 
\end{keywords}

\section{Introduction}

Our observational and theoretical knowledge about super-massive black holes (SMBHs), including the mechanisms by which the surrounding stars and gas interact with them, are continuously increasing (some examples are: \citealt{2002MNRAS.335..965Y, 2005MNRAS.363..223G, 2006ApJ...641..319P, 2009ApJ...698..198G, 2010MNRAS.407.1529H, 2012JPhCS.372a2041C, 2013arXiv1304.7762K}). Nuclear star clusters surrounding nuclear black holes have been observed over the last decade for a variety of galaxies \citep{2008ApJ...678..116S}. These clusters contain two distinct stellar populations that contribute significantly to their luminosities, one very old (several Gyr) and the other relatively young (a few to 100 Myr) indicating that star formation is ongoing in such systems \citep{2010IAUS..266...58B}. 

However, current instruments do not allow us to resolve the spatial distribution of young stars very close to the central black holes even for the  nearest (about 5-10~Mpc) extra-galactic nuclei. On the other hand, individual stars and young stellar associations can be studied in great detail for the Milky Way central region, at a distance of $\sim$~8~kpc from the Sun. The observed stars in the Galactic centre show a wide age range, in the nuclear cluster (including the S-stars) surrounding the SMBH and in at least one (possibly) warped disk further away, about 0.04-0.4~pc, from Sgr~A* \citep{2006ApJ...643.1011P, 2009A&A...499..483B, 2011ApJ...741..108P}. 

In addition to the disk stars and stellar members of the nuclear star cluster itself, \citet{Eckart04, 2013A&A...551A..18E}, \citet{2005A&A...443..163M} and \citet{muzic08} find that there are very young red objects, 0.1~pc away from Sgr~A*, in a compact group called IRS~13N. The authors propose that these are most probably Young Stellar Objects (YSOs) with 0.1-1~Myr age spread. The young age is inferred from their colors and dynamical considerations \citep{muzic08}. Remembering that the orbital period at about 1 pc (considering the inner edge of the circumnuclear disk (CND) and 
orbital eccentricities of molecular clumps) is about $10^4$~yr, this is an important evidence demonstrating that star formation in such a harsh environment close to a $4\times10^6 M_\odot$ black hole is probably in situ and frequent. It is worth to mention that IRS~13N is located (in projection) north of the IRS~13E complex. \citet{2005ApJ...625L.111S} and \citet{Fritz2010} show that IRS~13E group contains at least three blue/WR stars. The IRS~13E stellar sources are therefore more evolved stars than those belonging to IRS~13N. In fact, \citet{muzic08} show that the two groups are dynamically distinct based on proper motions. 

There are several numerical models attempting to explain the observed disk stars (\citealt{2005MNRAS.364L..23N, 2007MNRAS.379...21N, 2008Sci...321.1060B, 2012ApJ...749..168M}). In the mentioned models, infall of a giant molecular cloud is followed using  $\sim10^{4-5} M_{\odot}$ gas inside $\sim 1$~pc radius (15~pc in the case of \citealt{2012ApJ...749..168M}) or the evolution of a gaseous disk (initially surrounding the SMBH) is studied. While many properties of the observed young stars are addressed in these models, none of them could reproduce formation of compact stellar groups very close to Sgr~A*.

The focus of our current work is therefore to propose a possible scenario on the star formation in small systems very close to Sgr~A*. To reach this goal we study the difference between evolution of orbiting clumps and isolated clumps. We particularly compare our modelling with the IRS~13N YSO candidates observed in the vicinity of Sgr~A*. The importance of this modelling is to investigate star formation condition on a clump-scale within CND, as well as constraining the origin of the G2/DSO infrared source, as its highly eccentric orbit and possible stellar origin from the CND is not yet completely clear. 

We describe our modelling prescription in Section 2. We present our main results in Section 3 and compare clumps densities and the resulting protostars properties in different orbiting models. In Section 4, we discuss about collisions between clumps in the CND and also propose a scenario for the origin of the G2/DSO source. The conclusions of the paper are listed in Section 5. 

\section{Modelling Approach and AMUSE Application}

Motivated by the observations mentioned above, in particular \citet{Eckart04, 2013A&A...551A..18E} and \citet{muzic08}, we follow the evolution of a small molecular clump (MC) containing $100~M_\odot$ closely orbiting a $4\times10^6~M_\odot$ black hole including a simple prescription for star formation (described in detail in Section 2.1 and 2.2).  In this study, we mainly investigate if the interactions between the central SMBH and a small clump can lead the clump densities to higher values than the tidal density of the SMBH.

\begin{figure}
  \includegraphics[width=0.89\columnwidth]{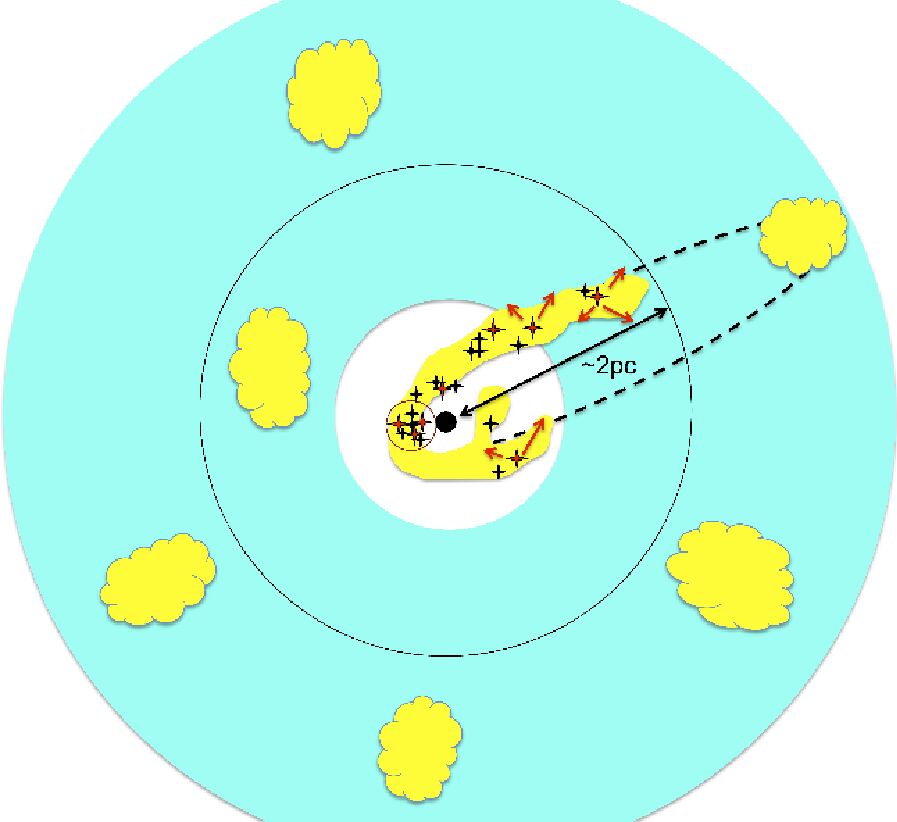}\hfill
  \caption{One modelled orbital set up is displayed for eccentricity~$\sim0.95$. The circumnuclear disk (cyan) is a clumpy (yellow) environment. Red (black) crosses represent massive (less massive) stars. The possible radiation from massive young stars is shown by red arrows. The thin red circle close to the black hole marks a IRS~13N-like configuration. The influence radius of Sgr~A* is shown by the black circle at $\sim$2~pc. More detailed structures, such as the disk stars, are not shown for clarity.}
\label{scheme}
\end{figure}

Figure~\ref{scheme} schematically demonstrates the idea we apply in the present work. This set up is also consistent with the observed configuration of the CND, the mini-spiral, IRS~13E and IRS~13N complexes, shown for example in figure 1.1 of \citet{2010RvMP...82.3121G}. This approach could be generalized for all the nucleated disk galaxies harboring a star forming molecular disk or ring.

To better visualize the observed central 5~pc of the Galaxy, it is helpful to remind that there are a few (neutral and ionized) streamers around Sgr~A* forming what is referred to as the mini-spiral. The mass of the ionized gas in the mini-spiral and neutral gas in the central parsec are estimated to be $\sim100$ and $300~M_\odot$, respectively (\citealt{1993ApJ...402..173J, 2005ApJ...622..346C}). Thus, MCs similar to those we model here are observed in the CND.

There are a few possibilities for the origin of the observed mini-spiral and the small stellar groups. One could be that a massive MC within the CND experiences encounters with other clumps, dissipates orbital energy during that encounter, loses angular momentum and consequently falls in toward the SMBH. We study a simplified version of this scenario. Another case could be that some small clumps separate from a giant molecular cloud while some regions of the parent giant cloud are about to be captured by the SMBH (a scenario often modelled to explain the observed young disk stars). It is also possible that a small MC is a byproduct of a collision between two large molecular clouds or a collision of one molecular cloud with a disk-like structure (the scenario recently modelled in \citealt{2013ApJ...771..119A}).

We model three families of MCs through this work: a) an isolated clump, b) an orbiting clump in a nearly circular orbit (eccentricity~$\sim$~0.5) and c) an orbiting clump in highly eccentric orbits (eccentricity~$\sim$~0.94). The choice of orbital parameters (peri- and apo-centre distances) of the two orbiting families of models are motivated by the size of the CND ($\sim$~1-5~pc). Isolated clumps are our reference models, to contrast our orbiting clumps about a SMBH with, and also to help compare our modelling prescription with other numerical studies.
 
We design the orbiting clumps mainly to study MC evolution and star formation as a function of different orbital configurations about a SMBH. We compare our results specifically with IRS~13N close to Sgr~A* if our current scenario could be one of its possible formation routes. 

We use smoothed particle hydrodynamics (SPH) method to simulate the evolution of a MC. We perform this modelling in the AMUSE\footnote{\texttt{www.amusecode.org}} (Astrophysical Multipurpose Software Environment) framework (\citealt{PortegiesZwart2009, 2013arXiv1307.3016P}). AMUSE is a python-based framework for astrophysical simulations, and allows combining the existing codes for stellar dynamics, stellar evolution, hydrodynamics and radiative transfer. The evolution of our model clump is followed using {\it Fi} \citep{2004A&A...422...55P, 2005PhDT........17P}, a ``community code" \citep{2013CoPhC.183..456P} integrated in AMUSE. The main physics in the current study is the self-gravity of the gas and the stellar interaction between newly formed stars.

\subsection{Clump Initial Conditions and Orbital Set up}

We model a spherical MC containing 100~$M_{\odot}$ within 0.376~pc diameter. To investigate star formation, these values are typical for a small molecular clump \citep[Table~1 in][on the cloud classification]{2007ARA&A..45..339B}. The clump initially has a uniform density with a random Gaussian divergence free velocity structure, with a power spectrum slope of $k=-3$ following \citet{2003MNRAS.343..413B}. 

We assume isothermal thermodynamics with mean molecular mass of 2.46. Two different initial temperatures are tested in our present simulations: 10 and 50~Kelvin. The higher temperature would be a better representative for the Galactic centre environment (\citealt{2013A&A...550A.135A} and references therein). We remind that our main goal is to probe whether our model clumps could survive in the strong tidal field of the SMBH close to the peri-centre  as the clumps are stretched along the orbits. To reach this goal, assuming isothermal gas is a good first approximation that allows us comparing clumps evolution on different orbits and exploring the role of the tidal field of the SMBH on small-scale model clumps. \citet{2012ApJ...749..168M} recently tested isothermal as well as radiative cooling thermodynamics for a similar but larger-scale problem. Their results show little difference between the two approaches in terms of stellar mass function.

We do not follow the higher densities including adiabatic regime. We therefore can not constrain the exact number of protostars as we do not resolve opacity limit. However, limiting our current work with the most straight forward thermodynamics (i.e., isothermal), we can investigate the general behaviour of model clumps depending on orbital configurations and compare the different orbiting models with the isolated models.

The detailed properties of early star formation require dedicated modelling, such as including a more realistic equation of state (e.g. polytropic) and resolving the opacity limit \citep[e.g. ][]{2007ApJ...656..959K, 2012MNRAS.419.3115B}.

Given our assumed clump mass and the number of particles, our mass resolution is 0.0192 $M_{\odot}$ (as we describe in detail in the next section) which is well below brown dwarf mass (0.08 $M_{\odot}$). So, we can model low to intermediate-mass protostars, typically more massive than 0.5-1~$M_{\odot}$ for our case. Clumps properties and orbital parameters are tabulated in Tables~\ref{table1} and~\ref{table2}. We include here a higher resolution run (using $ 1.0\times10^6$ particles) for the highly eccentric model to validate our results. We have also performed simulations with lower resolution, using 2.5$\times10^5$ particles, for all models to check our numerical results for convergence.

For the orbiting clumps, we design the orbits (varying eccentricity and peri-centre distance) such that one and a half orbital periods (second peri-centre passage) are about 1.25 times the initial free fall time ($t_{ff}$), $\sim$0.175~Myr. We initiate the orbiting clumps from the orbital apo-centre. The sound timescale of clumps is about 1~Myr. The sound speeds for the 10 and 50~Kelvin clumps are $\sim$~184 and 411 m/s, respectively. We integrate the clump evolution for 0.2 Myr. The minimum time step is chosen such that it can resolve the $t_{ff}$ at maximum density, $2-5\times10^{-5}$~Myr in practice. 

\begin{table*}
\begin{minipage}{158mm}
\centering
\caption{Overview of the clumps initial configuration.}
\label{table1}
\begin{tabular}{ccccccc}
\hline
Clump Mass & $N_{particles}$ & Clump Radius & Temperature & Initial Clump $t_{ff}$ & Accretion/Merging Distance & $\rho_{\mathrm{thresh}}$ [$\frac{amu}{cm^3}$]\\
\hline
\hline
100 [$M_\odot$] & 0.5$\times 10^6$ & 0.188 pc & 10 K & 0.135 Myr & 20 AU & 2.69 $\times10^{9}$ \\
100 [$M_\odot$] & 0.5$\times 10^6$ & 0.188 pc & 50 K & 0.135 Myr & 20 AU & 3.39 $\times10^{11}$\\

100 [$M_\odot$] & 1.0$\times 10^6$ \footnote{\text{This additional run has been performed with higher resolution to numerically check the highly eccentric model}} & 0.188 pc & 10 K & 0.135 Myr & 20 AU & 1.05 $\times10^{10}$\\

\hline
\end{tabular}
\end{minipage}
\end{table*}

\begin{table*}
\begin{minipage}{158mm}
\centering
 \caption{The orbital parameters for orbiting models.}
\label{table2}
\begin{tabular}{cccccc}
\hline
Semi-major Axis & Eccentricity & Peri-centre & Apo-centre & Orbital Period & 2nd Peri-Passage time\\
\hline
\hline
1.8 pc & 0.944 & 0.1 pc & 3.5 pc & 0.113 Myr & 0.17 Myr\\
1.8 pc & 0.5   & 0.9 pc & 2.7 pc & 0.113 Myr & 0.17 Myr\\
\hline
\end{tabular}
\end{minipage}
\end{table*}

\subsection{SMBH and Sink Description}

Orbiting clumps are set to follow a Keplerian orbit about a $4\times10^6 M_{\odot}$ black hole close to the mass of Sgr A*
\citep{2008ApJ...689.1044G, 2009A&A...502...91S, 2009ApJ...692.1075G}. Note that the influence radius of such a black hole is about 2~pc \citep{2005PhR...419...65A}. We consider a point potential to model the SMBH and exclude the possible effect of the nuclear star cluster on the clump evolution. 
The gravitational softening length in our models is fixed to $r_{\mathrm{peri-centre}}/$5, ranging from 0.02 to 0.2~pc depending on the orbital configurations. 

Protostars are modelled using a sink approach (\citealt{1995MNRAS.277..362B, 2003MNRAS.343..413B} and references therein) in such a way that gas particles with densities higher than a certain threshold, $\rho_{\mathrm{\mathrm{thresh}}}$, are replaced by point mass particles (sinks) that only interact with the remaining particles via gravity. The $\rho_{\mathrm{thresh}}$ is set by the requirement for self gravitating SPH simulations of resolving the local Jeans mass with at least $1.5 \times N_{\mathrm{neighbours}} = 96$ particles. The importance of this point is discussed in \citet{1997MNRAS.288.1060B}. 

The initial clump density is 1.45$\times10^{5}$ amu$/$cm$^{3}$ and $\rho_{\mathrm{thresh}}$ values are a few $10^{9}$ and a few $10^{11}$~amu$/$cm$^{3}$ for 10 and 50 Kelvin models, respectively. Our $\rho_{\mathrm{thresh}}$ values are higher than tidal density of the SMBH, $10^{8}$~amu$/$cm$^{3}$
 at 1 pc where the clump spends more time in that region, and about $10^{11}$~amu$/$cm$^{3}$ at about 0.1 pc (\citealt{2013ApJ...767L..32Y}). Sinks will be replaced unless there is already a sink particle within the smoothing kernel of the particle ($\sim$10-50~AU for 50 and 10~Kelvin models). We also allow sinks to merge with each other if their distance is less than 20~AU. Sink particles also accrete any dense gas particle within this distance. Thus, for a 100 $M_{\odot}$ clump using 0.5$\times10^6$ particles, we interpret the sink particles with a mass larger than 0.0192~$M_{\odot}$ (or $\sim$~20~$M_{\mathrm{Jupiter}}$) as being protostellar objects, but note that we do not follow any additional fragmentation below our resolution of $\sim10$ AU.

For the scope of this paper, we do not include radiative transfer in our clump evolution. We also do not include stellar feedback in any format (mechanical as well as radiative). Excluding stellar feedback could be a good simplification as we stop integrating the simulations as soon as stars form.

Our chosen clump mass and size as well as our sink accretion threshold distance (20~AU) are conservative assumptions to reproduce IRS~13N-like groups and stellar mass distribution. In other words, choosing larger and more massive clumps, together with larger accretion distance criterion for sinks growth can lead to the formation of higher number of protostars, and a mass function with top-heavy IMF trend.

\section{Results}

\begin{figure*}
\includegraphics[width=0.65\textwidth]{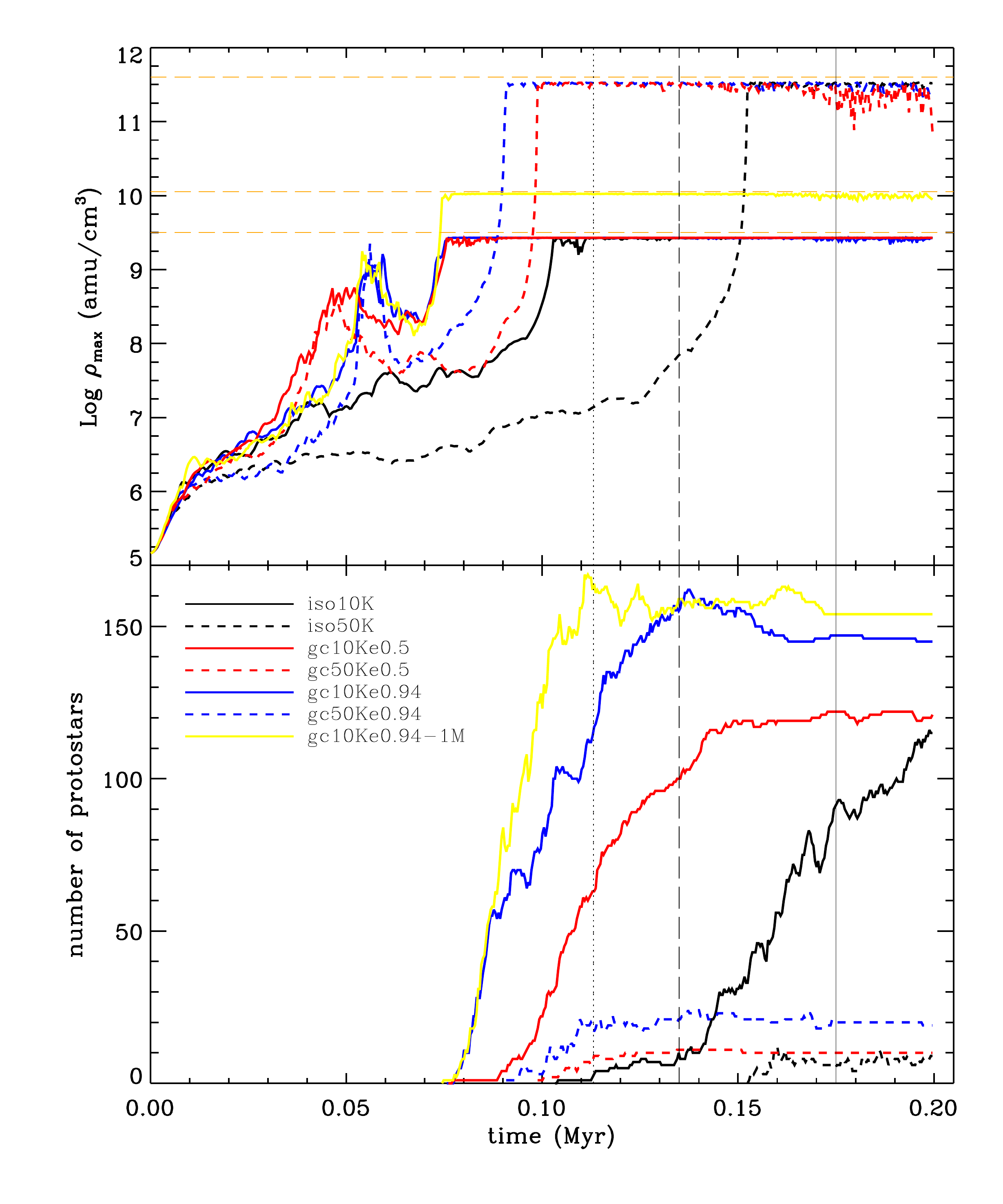}
\caption{The maximum gas densities and the number of protostars for each model is plotted as a function of time from the beginning of simulations. Symbols of each models are consistent in both panels. One orbital period is 0.113~Myr (vertical dotted line), one free fall time is 0.135~Myr (vertical dashed line) and the second peri-centre passage is 0.175~Myr (vertical solid line). Orange horizontal lines are $\rho_{\mathrm{thresh}}$ values for 10 and 50 Kelvin models, respectively at about 2.69~$\times10^{9}$ and 3.39~$\times10^{11}~\frac{amu}{cm^3}$. [see the online version for full color information]}
\label{gas_sink}
\end{figure*}

\begin{figure*}
\includegraphics[width=0.65\textwidth]{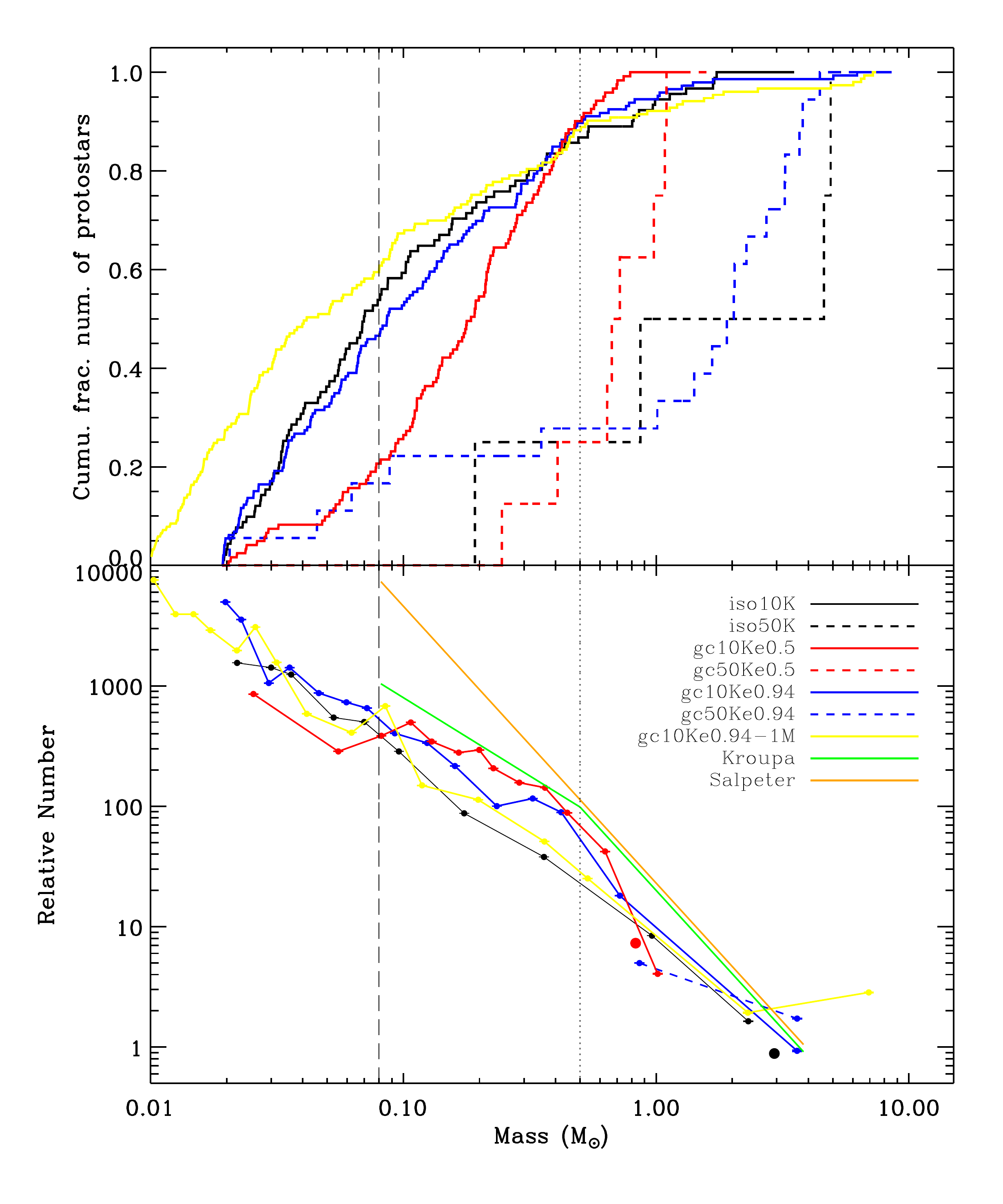}
\caption{The cumulative number of protostars and the IMF of all models computed at 1.3~$\times~t_{ff}\sim$0.175 Myr. This is about 1.55 times the period of orbiting models. The black solid line represents the 10 K isolated model (our reference model). Symbols of all the models for both panels are the same. In the lower panel, there is only one data point for the iso50K (larger black dot) and gc50Ke0.5 (larger red dot) models as there are only a few stars. The Kroupa and Salpeter slopes are overplotted with green and orange lines, respectively. Vertical lines are 0.08~$M_{\odot}$ (brown dwarf) and 0.5~$M_{\odot}$ limit. [see the online version for full color information]}
\label{imf100msun}
\end{figure*}

\begin{table*}
\begin{minipage}{168mm} 
\centering
\caption{Summary of star formation properties for 100~$M_\odot$ clumps, computed at 1.3 $\times~t_{ff}$.}
\label{table3}
\begin{tabular}{ccccccc}
\hline
Model Clump & $N_{stars}$  & $N_{BDs}$ & Mean Mass [$M_\odot$] & Maximum Stellar Mass [$M_\odot$] & Total resolved Stellar Mass [$M_\odot$] & SFE\\ 
\hline
\hline
iso10K & $\geq$92-49 & 49 & 0.27 & 3.51 & 24.9  & 25$\%$\\ 
iso50K & $\geq$5     & 0  & 3.49 & 6.91 & 17.5  & 18$\%$\\
\hline
gc10Ke0.5 & $\geq$122-25  & 25 & 0.24   & 1.24 & 29.18 & 29$\%$\\ 
gc50Ke0.5 & $\geq$9       & 0  & 0.83   & 1.62 & 7.46  & 7$\%$\\
\hline
gc10Ke0.944 & $\geq$147-68 & 68 & 0.33 & 8.56 & 48.8  & 49$\%$\\
gc50Ke0.944 & $\geq$19-3   & 3  & 2.2  & 7.83 & 41.87 & 42$\%$\\
gc10Ke0.944-1M \footnote{\text{The additional run with higher resolution to numerically check the highly eccentric model}} & $\geq$154-91 & 91 & 0.43 & 7.42 &  66.16 & 66$\%$\\
\hline
\end{tabular}
\end{minipage}
\end{table*}

\subsection{Qualitative Description}

As the upper panel of Figure~\ref{gas_sink} shows, the maximum gas densities in all models (except the iso50K model) reach the corresponding $\rho_{\mathrm{thresh}}$ values earlier than one free fall time (vertical dashed line at 0.135~Myr). However, for all the orbiting models this happens even earlier than in the case of their counterpart isolated models. The required $\rho_{\mathrm{thresh}}$ is reached at around half of the orbital period when the clump is experiencing the first peri-centre passage, feeling the orbital compression (see Appendix). This suggests that the orbital compression of a MC close to the SMBH acts as an external pressure on the clump, and facilitates the formation of stars for the high eccentricity clumps. It is known that the gas densities should be above $\sim10^8$~cm$^{-3}$ in order for star formation to occur close (within the central 1 pc) to Sgr~A* (e.g. \citealt{2013ApJ...767L..32Y}). As it can be seen in Fig.~\ref{gas_sink}, this is in fact the case in our orbiting models, while the initial clump density is just about $10^5~$cm$^{-3}$.

The maximum gas density in gc10Ke0.5 model (red solid line in Fig.~\ref{gas_sink}) reaches its $\rho_{\mathrm{thresh}}$ slightly after the first peri-passage, much earlier than gc50Ke0.5 model (red dashed line). The gc10Ke0.5 clump is constantly compressed to form protostars as the Jeans mass in the 10~K model is lower ($\sim$~1~$M_{\odot}$) than in the case of the 50~K model ($\sim$~10~$M_{\odot}$). For the highly eccentric orbit (e$=$0.944), in both the 10 and 50 K models, the relevant $\rho_{\mathrm{thresh}}$ limits are reached just at the half-period and protostars start forming right after the first peri-centre passage. This shows that the SMBH helps to orbitally compress the gas, and the densities rise to the necessary $\rho_{\mathrm{thresh}}$ quicker than in the case of isolated clumps or the models with the milder eccentricity. The SMBH plays a supportive role in star formation by causing the the gas densities increase towards the threshold values during the first peri-centre passage. 

Figure~\ref{gas_sink} lower panel, presents the number of formed protostars for each model in terms of time. The stellar numbers are consistent with  the gas density evolution. In some models, protostars with a mass above the resolution, $\sim0.0192~M_{\odot}$, appear with a short delay after the threshold densities have been reached. In gc50Ke0.944 model, protostars can still form but are less numerous and more massive compared with the gc10Ke0.944 model because of the higher Jeans masses in the 50~K models. The number of stars in $e=0.5$ models are between the isolated and the high eccentric ($e=0.944$) ones.

The number of protostars varies slightly over time due to merging of sinks or scattering and possibly being ejected from the clump. This effect is more prominent in the highly eccentric models (blue lines) where the clump is more orbitally compressed and the number density of stars inside the clump is higher, consequently stellar encounters as well as merger possibilities are more frequent.

The column density maps of the models at 0.7 to 1.4 times of the initial free fall time are illustrated in Figures~\ref{iso10K} to~\ref{gc50Ke0.5}.

\begin{table*}
\begin{minipage}{158mm}
\centering
\caption{The properties of IRS~13N-like groups in models with e=0.944.}
\label{table4}
\begin{tabular}{c c c c}
\hline
Model Clump & Time [Myr] & Individual Stellar Mass [$M_{\odot}$] & (x,y,z) [pc] from the SMBH\\
\hline
\hline
gc10Ke0.944 & 0.163 & 8.39, 1.26, 5.02, 1.39 & (-0.09866, -0.07554, -0.0041)\\
gc10Ke0.944 & 0.175 & 6.23, 0.76, 1.03, 1.02 & (-0.10429, -0.0927,  0.00015)\\
\hline
gc50Ke0.944 & 0.162 & 3.67, 3.20 & (-0.03905, 0.11998, 0.00028)\\
gc50Ke0.944 & 0.164 & 2.72, 4.42 & (-0.09221, 0.02037, -0.00038)\\
gc50Ke0.944 & 0.171 & 7.80, 2.01, 0.99 & (0.01041, 0.20435, 0.00158)\\
\hline
gc10Ke0.944-1M & 0.160 & 6.00, 2.52 & (-0.07851, -0.06054, -0.00092)\\
gc10Ke0.944-1M & 0.170 & 0.84, 1.54, 2.63, 3.76  & (-0.09429, 0.01838, -0.00043)\\
gc10Ke0.944-1M & 0.173 & 7.42, 7.19 & (-0.10787, 0.05193, -0.00141)\\
\hline
\end{tabular}
\end{minipage}
\end{table*}

\subsection{Stellar Mass Distributions: Isolated versus Orbiting Models}

We compute the initial mass function (IMF) and cumulative number of protostars at $1.3\times t_{ff}$, that is about 0.175~Myr for all the models and is equivalent to $\sim$1.55 times the orbital period for orbiting clumps (right after the second peri-centre passage).

In Figure~\ref{imf100msun}, the top panel depicts the cumulative number of protostars. There are generally fewer stars for the 50~K models (dashed lines) than in the 10~K models. The higher Jeans mass ($\sim$10~$M_{\odot}$) in the case of 50~K clumps prevents it from collapse and form stars as easily as in the case of 10~K models. However, the few formed protostars in the higher temperature models are normally massive, as can be seen in the dashed concaveness-shape curves in the top panel of Figure~\ref{imf100msun}. 

We note that our cumulative fractional number of stars for the isolated 10 K model is consistent with the one of \citet{2009MNRAS.392.1363B} (although he uses a barotropic equation of state for a 50~$M_{\odot}$ clump). This suggests that any difference between the results of our models is due to the presence of the SMBH and the different orbital configurations.

The statistical properties of protostars for different models are listed in Table~\ref{table3} and summarize the above results. We consider every object more massive than our mass resolution and below 0.08~$M_{\odot}$ as brown dwarfs and planet-size objects.

The mass distribution of protostars, i.e. initial mass function (IMF), is displayed in the lower panel of Figure~\ref{imf100msun}. The Kroupa \citep{2001MNRAS.322..231K} and Salpeter \citep{1955ApJ...121..161S} slopes are also plotted for comparison. This panel shows that between 0.08 to 0.5~$M_{\odot}$ the IMF of all models are relatively comparable to each other, and seem to be consistent (in this mass range) with the Kroupa slope. 
There is only one data point representing the IMF of the orbiting iso50K model as well as the gc50Ke0.5 model (one larger black and one larger red dot respectively), as there are less than 10 stars forming in these models. The gc50Ke0.944 model shows a top-heavy IMF trend that is also obvious from the concaveness of the dashed blue line in the upper panel of this figure. As we also mentioned in Section 2.1, we can not constrain the exact number of stars especially in the low-mass end, below 0.08~$M_{\odot}$, in which there are observational evidences for a decline in this range. Our current models cannot reproduce this behaviour as our star formation modelling does not include the opacity limit and we run relatively low resolution simulations, but note that this mass range is not the main focus of the current study.
 
The results of the higher resolution run (using $ 1.0\times10^6$ particles) in terms of number of stars and mass distribution are 
consistent with the main models (Table~\ref{table3}), however, its total stellar mass is larger than the corresponding 10 K model. Currently we cannot
exclude that this effect is linked to our simplified sink particle prescription described in Section 2.

It is worth noting that we have low number statistics in the current set of simulations, i.e. about 100 stars in each model and masses are non-uniformly distributed over a range of $\sim$0.01 to 10~$M_{\odot}$. Moreover, we focus on the formation of a relatively small group of (massive) stars. Therefore, care should be taken on interpreting and comparing IMF slopes and mass function trends. To compute the mass function we use adaptive binning such that there is always a fixed number (10) of stars in each bin. So, the relative uncertainty for each data point is constant and is $\sim$0.3.

\begin{figure*}
\centering
\resizebox{0.65\textwidth}{!}{
\vspace{-10pt}
\includegraphics[width=0.5\textwidth]{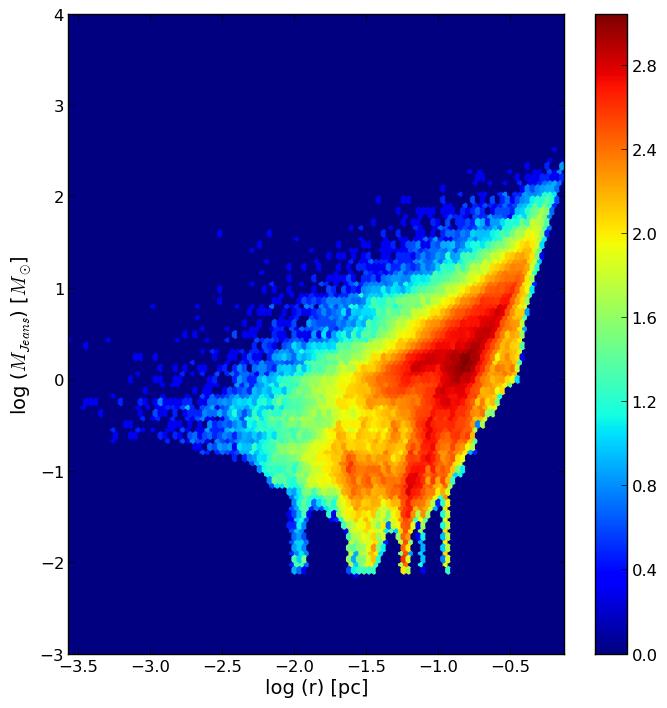}
\includegraphics[width=0.5\textwidth]{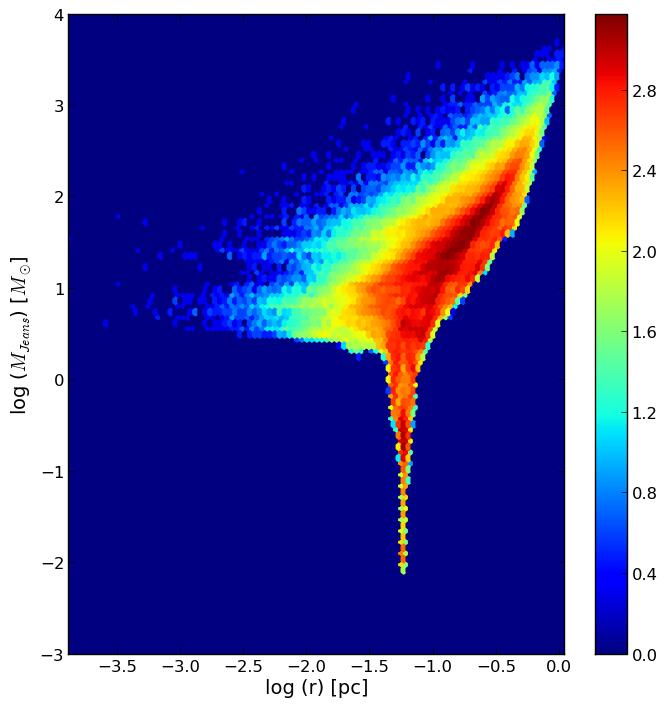}
}
\resizebox{0.65\textwidth}{!}{
\vspace{-10pt}
\includegraphics[width=0.5\textwidth]{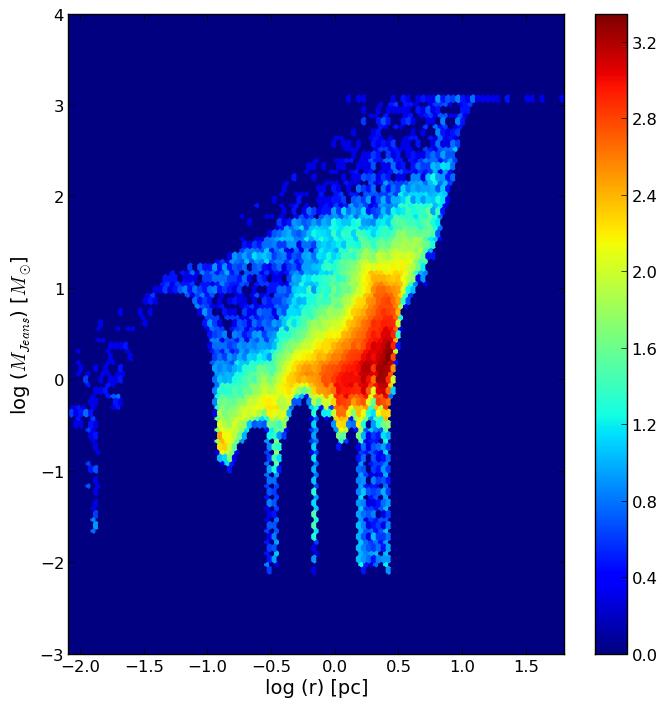}
\includegraphics[width=0.5\textwidth]{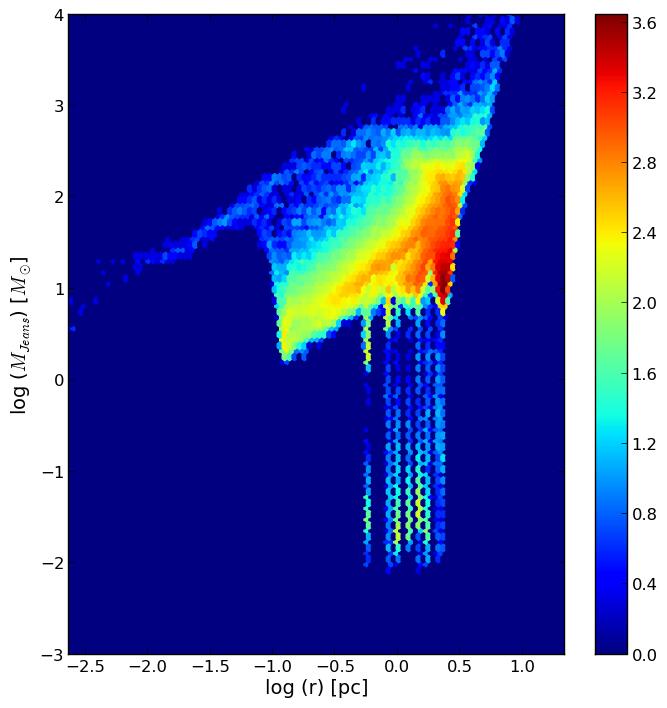}
}
\resizebox{0.65\textwidth}{!}{
\vspace{-10pt}
\includegraphics[width=0.5\textwidth]{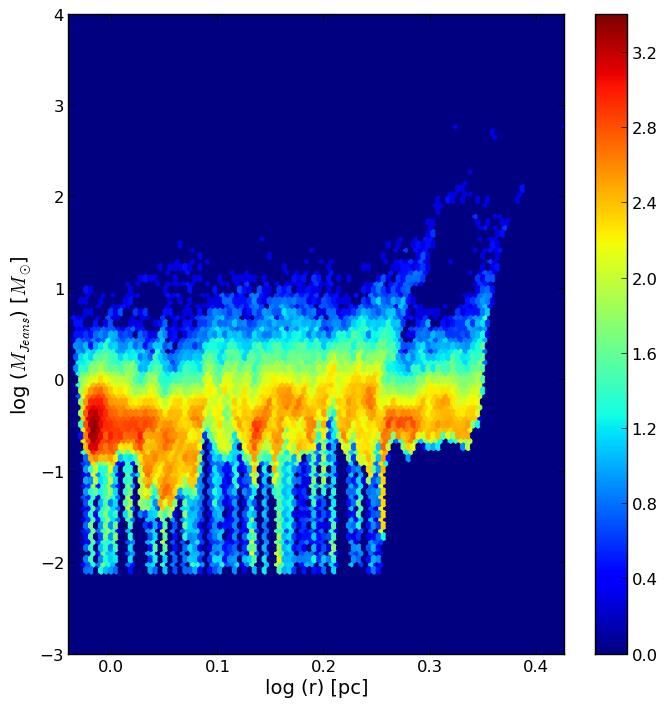}
\includegraphics[width=0.5\textwidth]{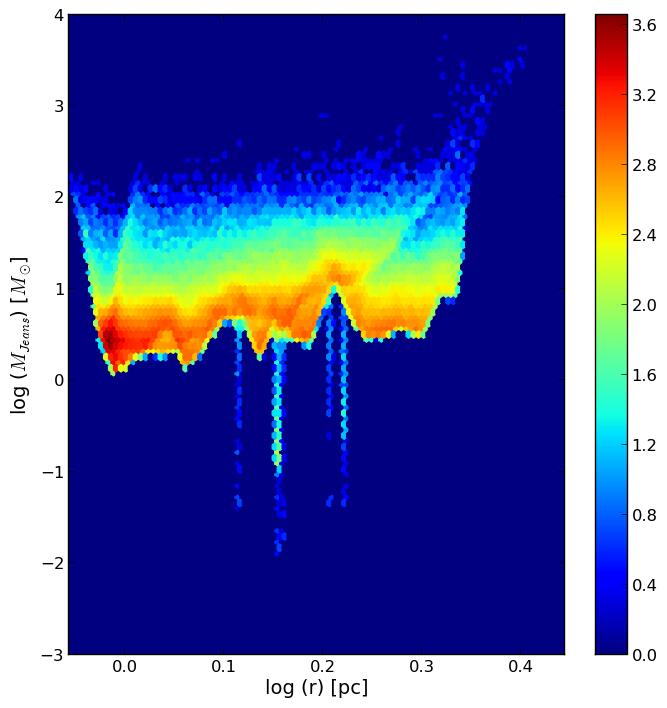}
}
\vspace{0pt}
\caption{The Jeans mass number density of gas particles versus distance to the SMBH (centre of the clump in the isolated models) is plotted at $1.3\times t_{ff}$ (0.175~Myr). The number density of gas particles decrease logarithmically from red to blue as shown with the colour bar. Top to bottom correspond to isolated, orbiting e0.944 and e0.5 models. The higher Jeans mass in the case of 50~Kelvin models (right column) explain the lower number of protostars forming in these models. Note that in the Jeans mass distribution of the high eccentricity models (bottom row) orbital compression of the clump (versus iso50K model) results in the formation of more stars.}
\label{jeans}
\end{figure*}

\begin{figure*}
\centering
\resizebox{0.65\textwidth}{!}{
\vspace{-30pt}  
\includegraphics[width=0.5\textwidth]{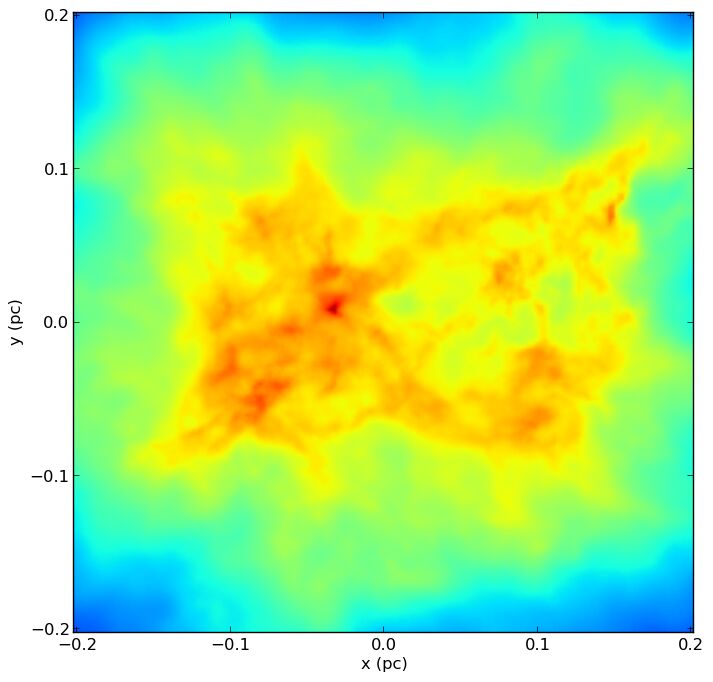} 
\includegraphics[width=0.5\textwidth]{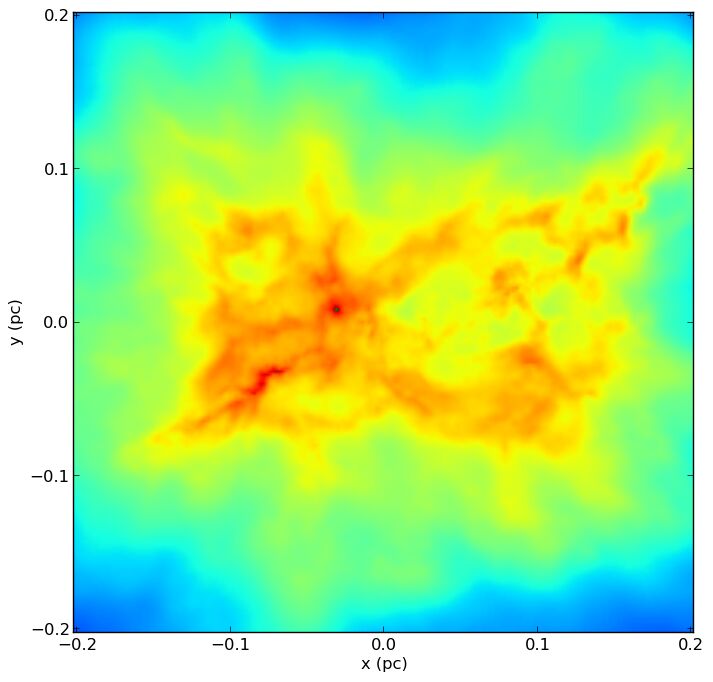} 
}
\resizebox{0.65\textwidth}{!}{
\vspace{-3pt}  
\includegraphics[width=0.5\textwidth]{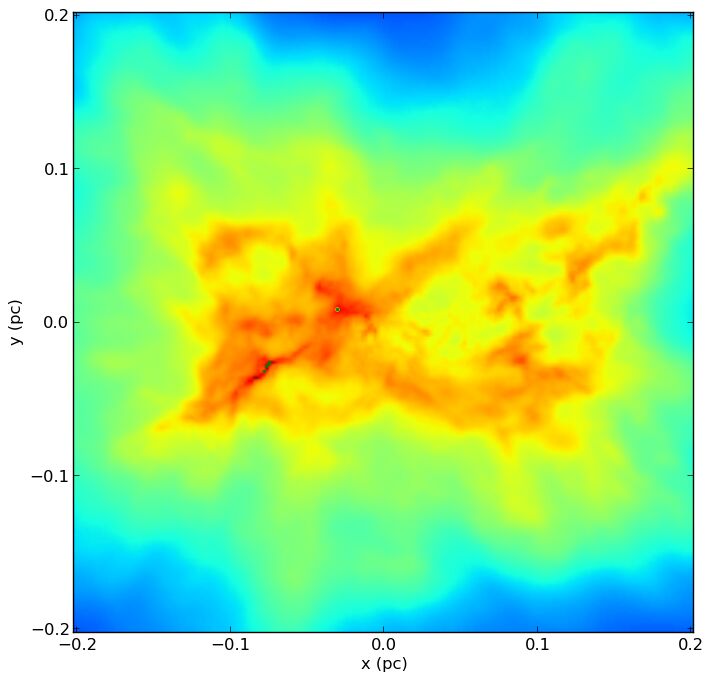} 
\includegraphics[width=0.5\textwidth]{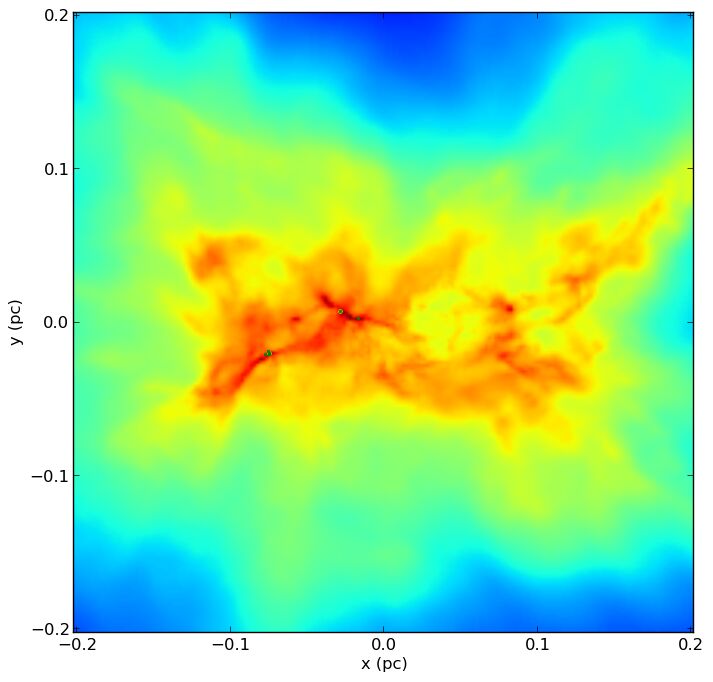} 
}
\resizebox{0.65\textwidth}{!}{
\vspace{-3pt}
\includegraphics[width=0.5\textwidth]{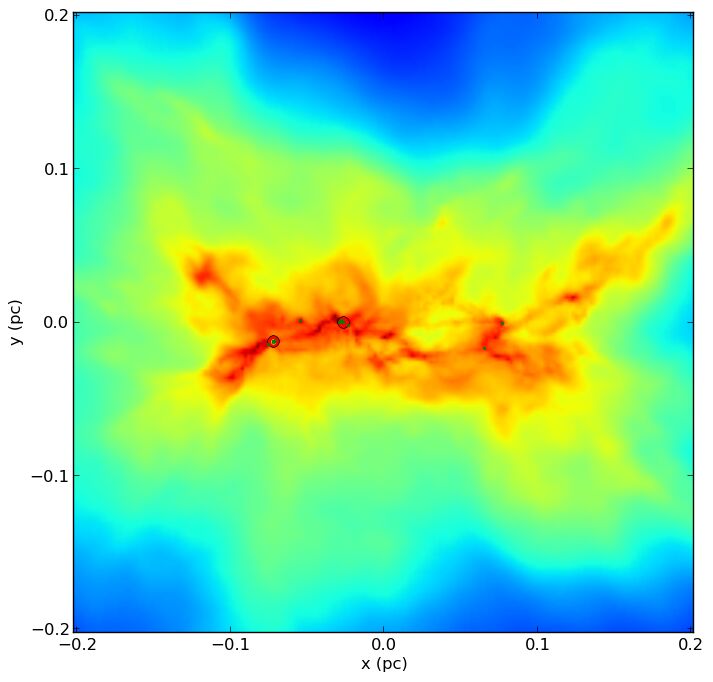} 
\includegraphics[width=0.5\textwidth]{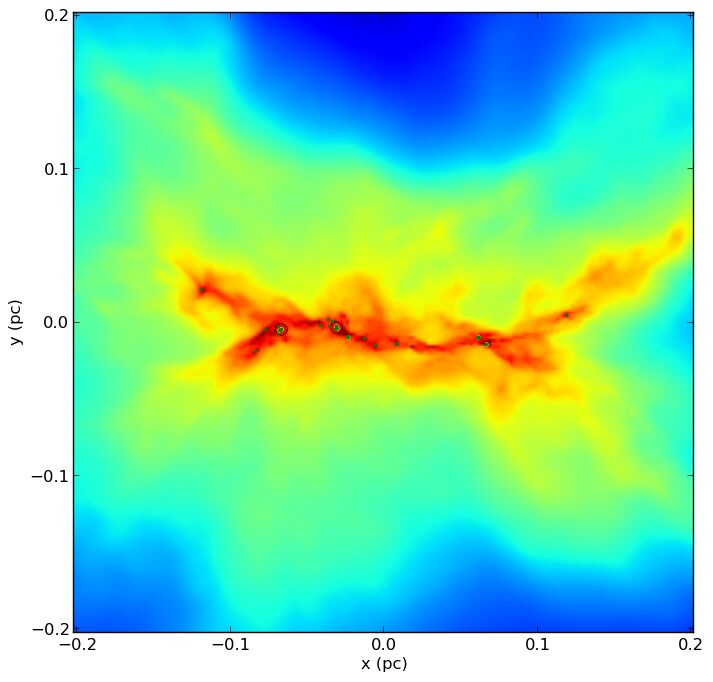} 
}
\resizebox{0.65\textwidth}{!}{
\vspace{-3pt}
\includegraphics[width=0.5\textwidth]{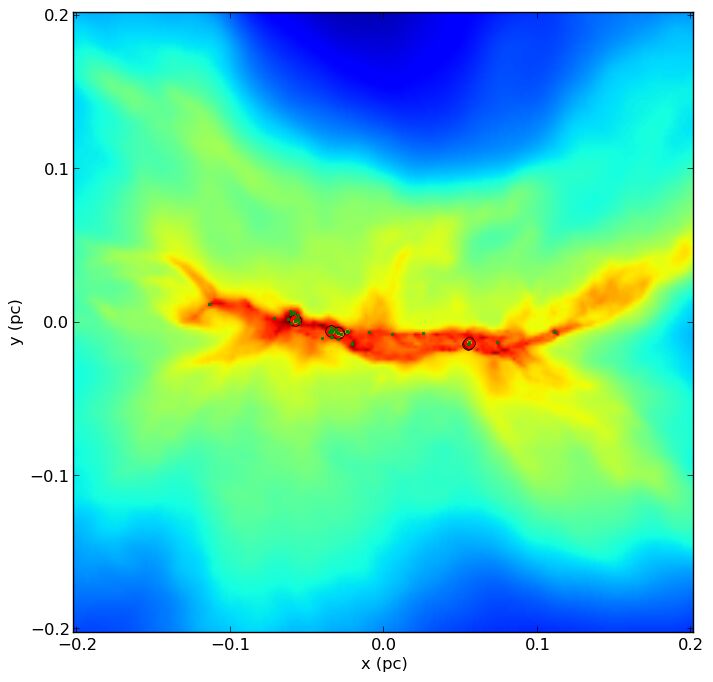} 
\includegraphics[width=0.5\textwidth]{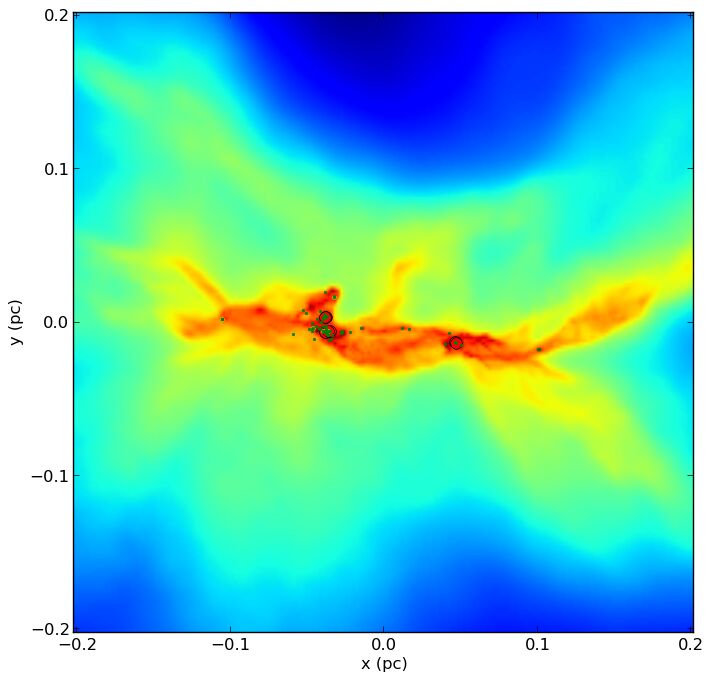} 
}
\vspace{-5pt}
\caption{The temporal evolution of column density map for iso10K model. Top left to bottom right correspond to 0.7 to 1.4 times the initial free fall time,~$\sim$0.095 to 0.19 Myr. The column densities span logarithmically over $\sim$20.0 to 24.5~$\frac{amu}{cm^{2}}$ increasing from blue to dark orange. The protostars are also overplotted on the map and marked with red, yellow and green dots, respectively for more massive than 1~$M_{\odot}$, more massive than 0.5~$M_{\odot}$ but less than 1~$M_{\odot}$, and less massive than 0.5~$M_{\odot}$. The formation of stars in two compact cores formed at about 1~$\times~t_{ff}$ is noticeable.}
\label{iso10K}
\end{figure*}

\begin{figure*}
\centering
\resizebox{0.65\textwidth}{!}{
\vspace{-30pt}
\includegraphics[width=0.5\textwidth]{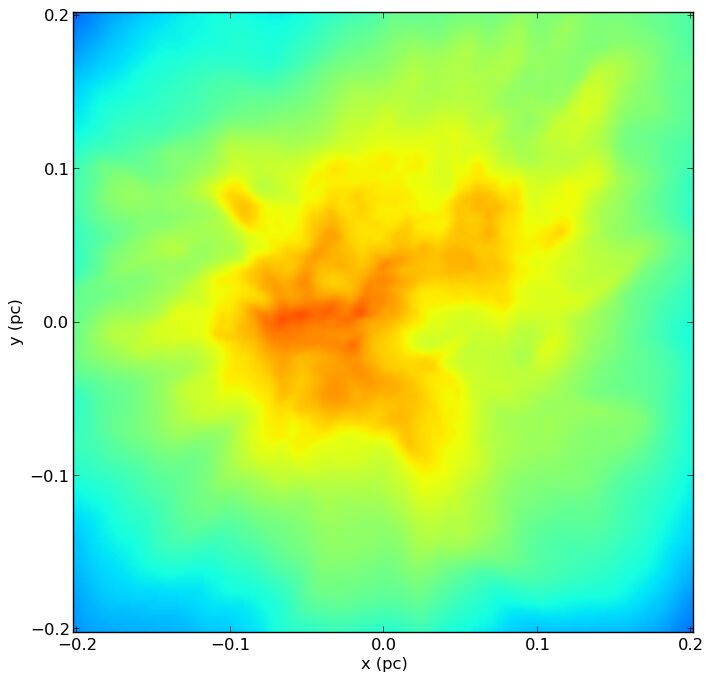}
\includegraphics[width=0.5\textwidth]{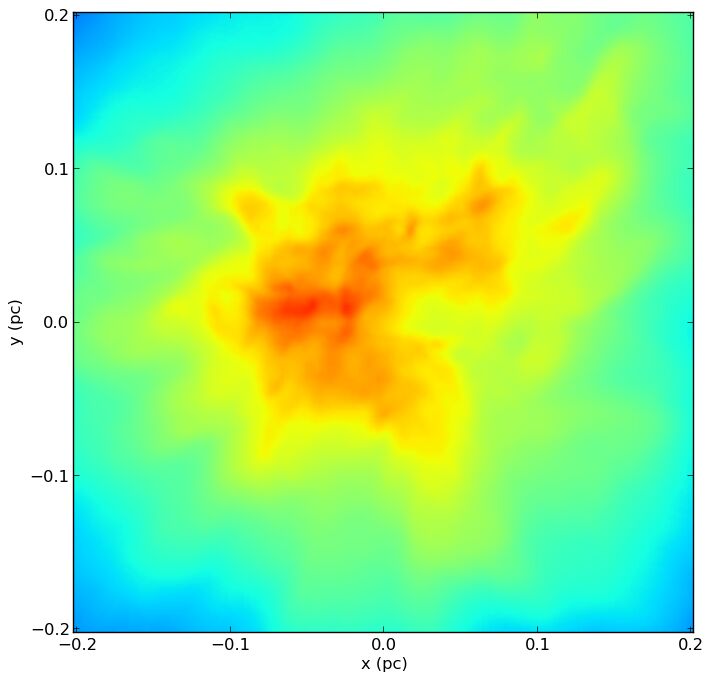}
}
\resizebox{0.65\textwidth}{!}{
\vspace{-3pt}
\includegraphics[width=0.5\textwidth]{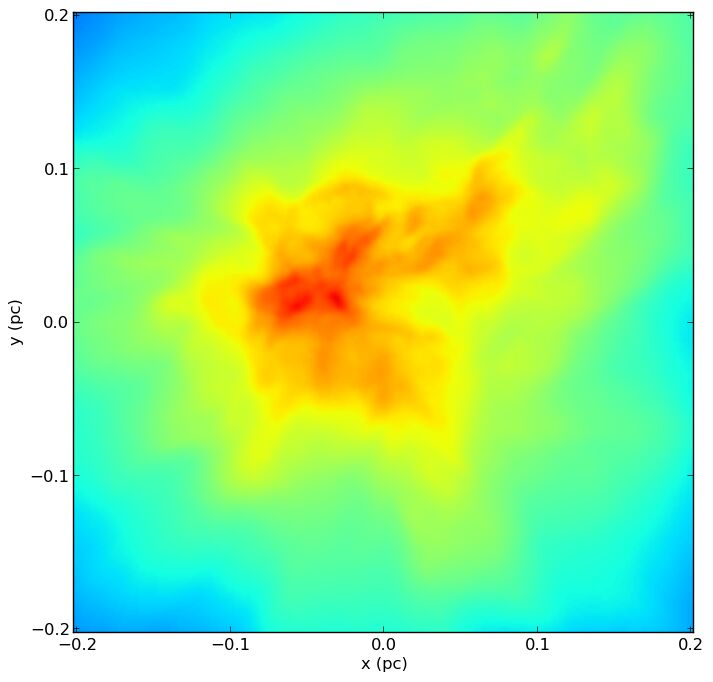}
\includegraphics[width=0.5\textwidth]{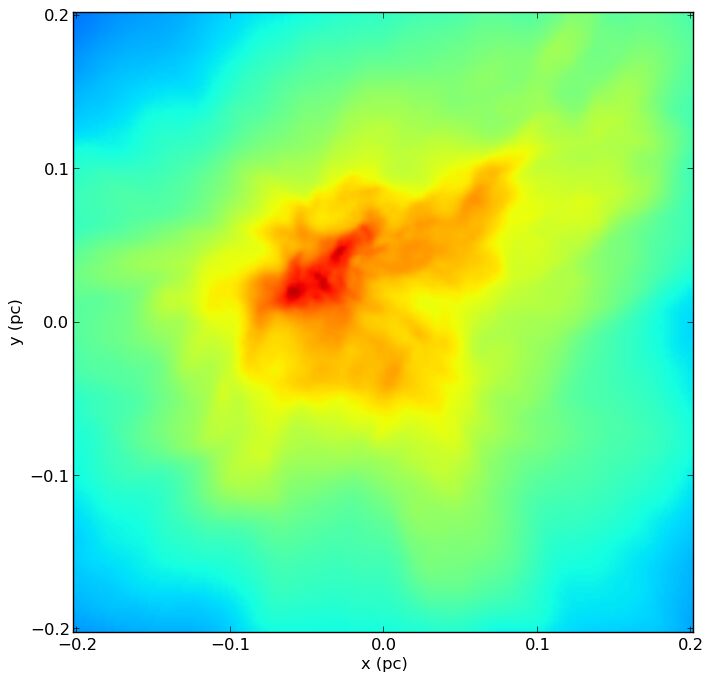}
}
\resizebox{0.65\textwidth}{!}{
\vspace{-3pt}
\includegraphics[width=0.5\textwidth]{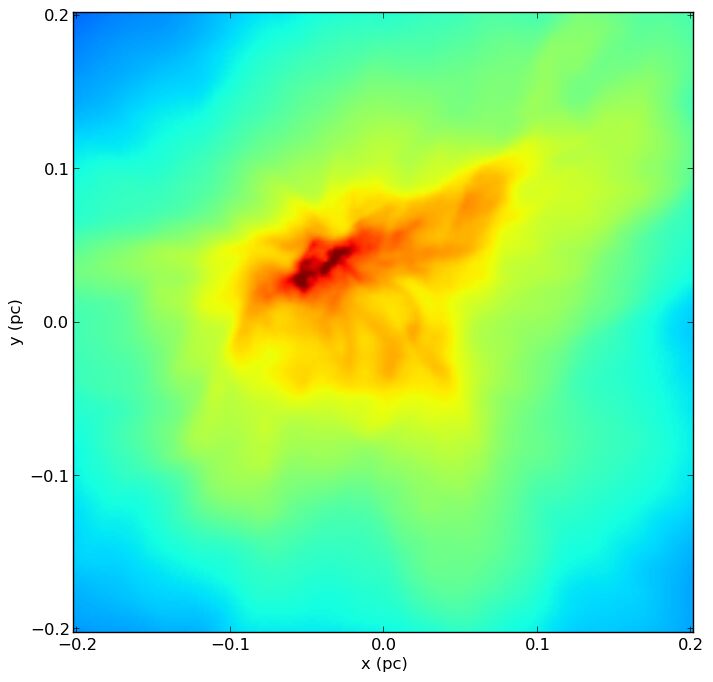}
\includegraphics[width=0.5\textwidth]{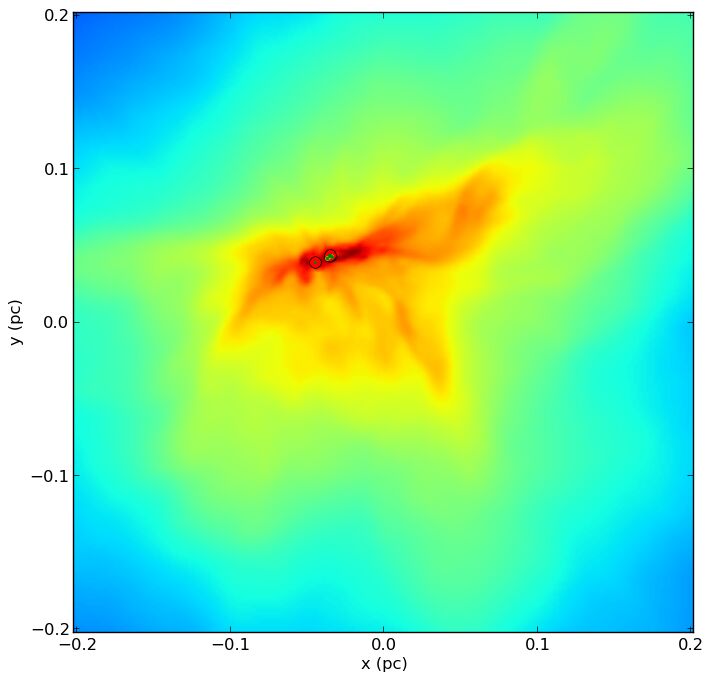}
}
\resizebox{0.65\textwidth}{!}{
\vspace{-3pt}
\includegraphics[width=0.5\textwidth]{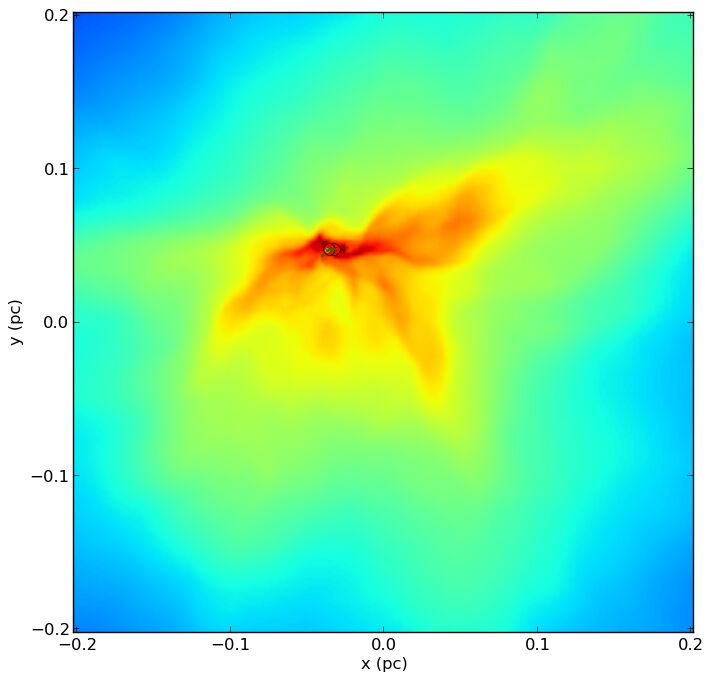}
\includegraphics[width=0.5\textwidth]{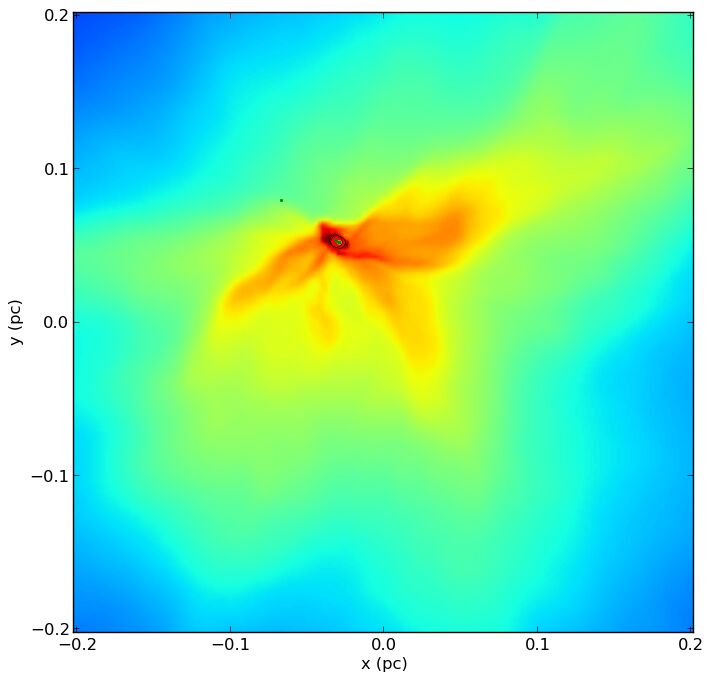}
}
\vspace{-5pt}
\caption{The temporal evolution of column density map for iso50K model. The temporal snaps, color code of densities and protostars symbols are as in the previous figure. The column densities span logarithmically over $\sim$21.0 to 25.0~$\frac{amu}{cm^{2}}$. There is a compact dense core forming in which less numerous protostars form than in the case of 10~K clump as the Jeans mass here is higher.}
\label{iso50K}
\end{figure*}

\begin{figure*}
\centering
\resizebox{0.65\textwidth}{!}{
\vspace{-30pt}
\includegraphics[width=0.5\textwidth]{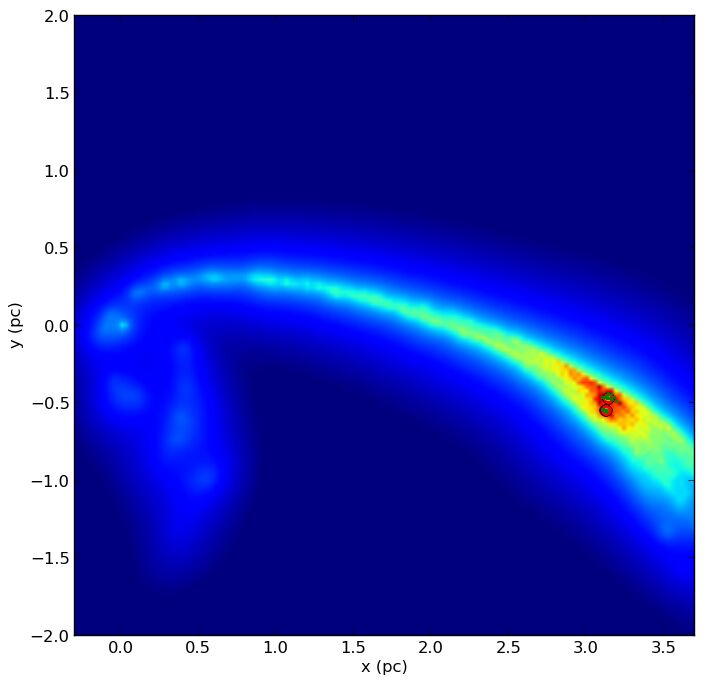}
\includegraphics[width=0.5\textwidth]{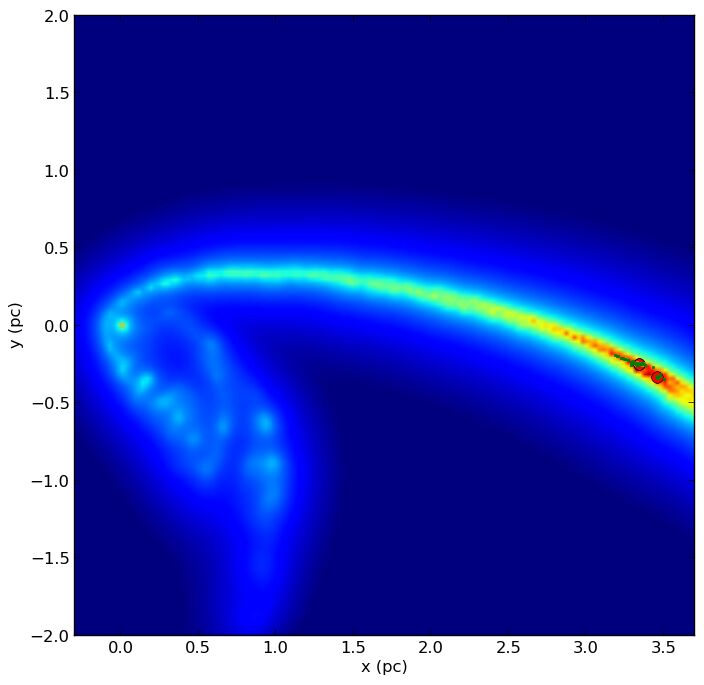}
}
\resizebox{0.65\textwidth}{!}{
\vspace{-3pt}
\includegraphics[width=0.5\textwidth]{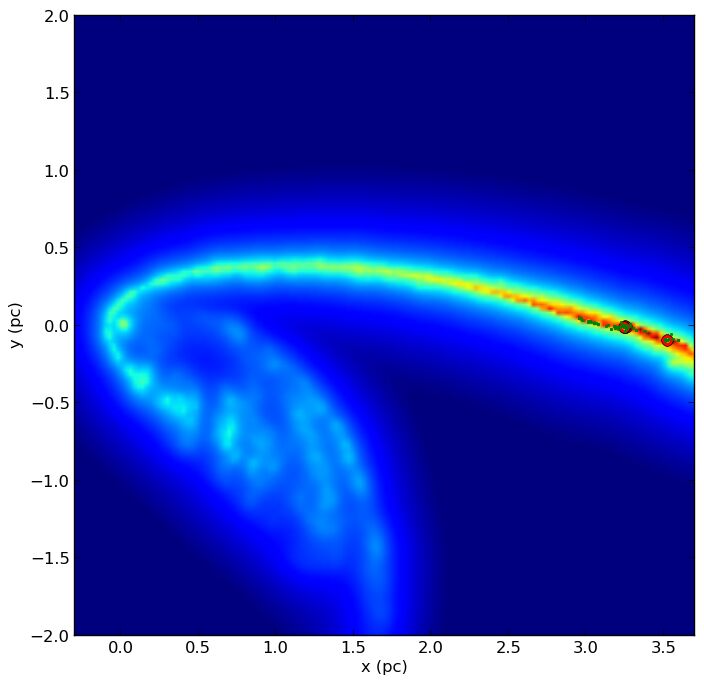}
\includegraphics[width=0.5\textwidth]{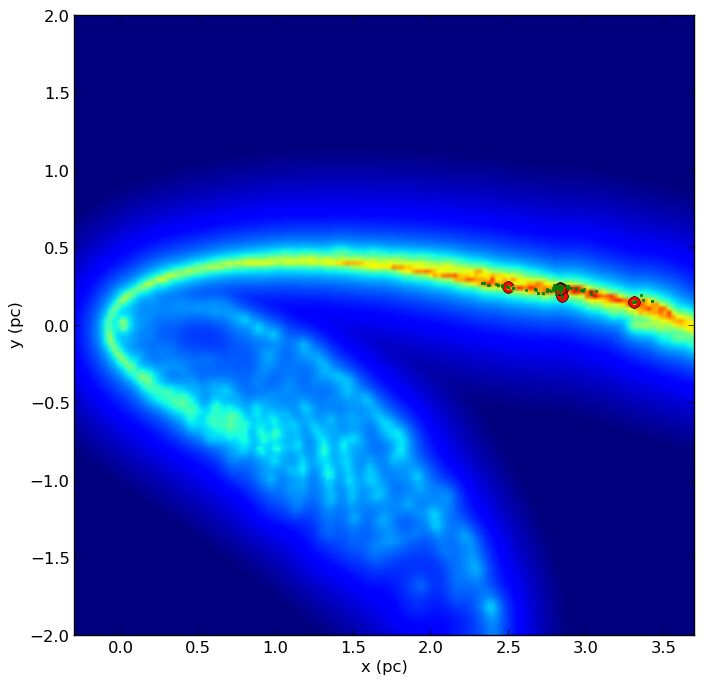}
}
\resizebox{0.65\textwidth}{!}{
\vspace{-3pt}
\includegraphics[width=0.5\textwidth]{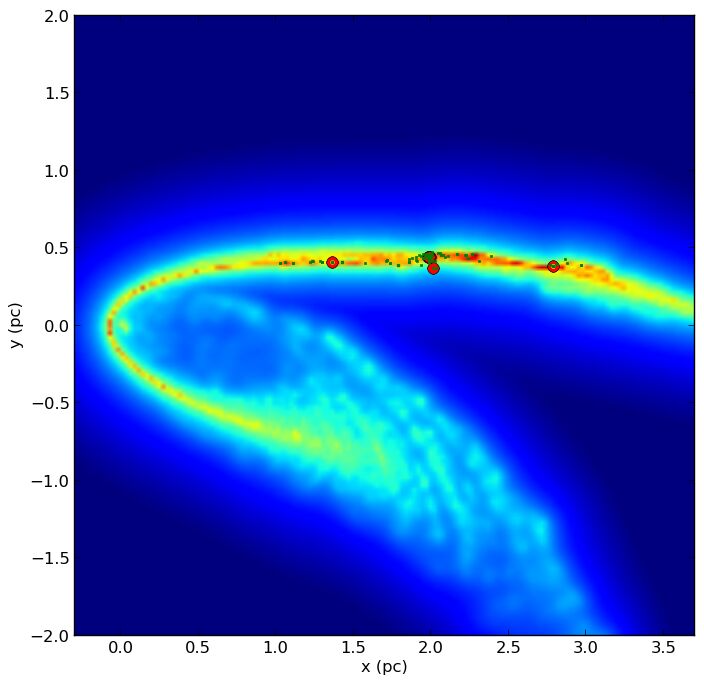}
\includegraphics[width=0.5\textwidth]{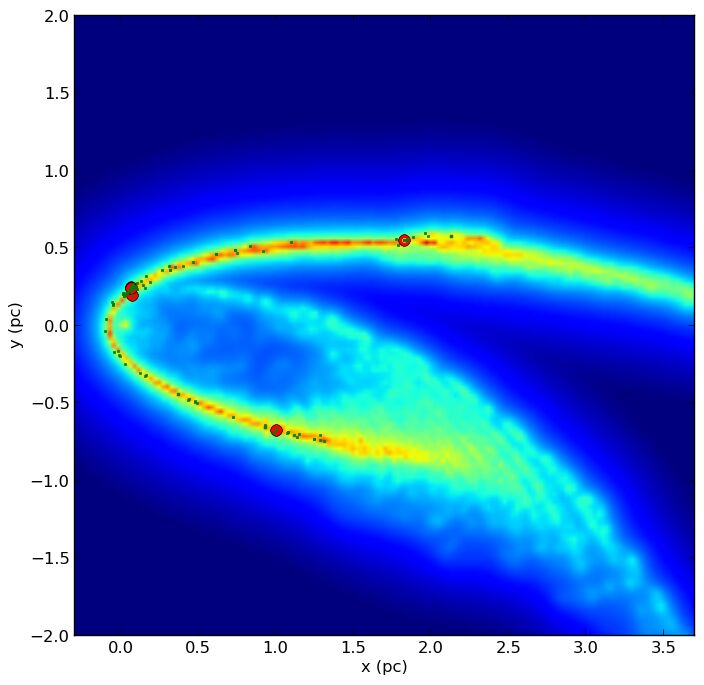}
}
\resizebox{0.65\textwidth}{!}{
\vspace{-3pt}
\includegraphics[width=0.5\textwidth]{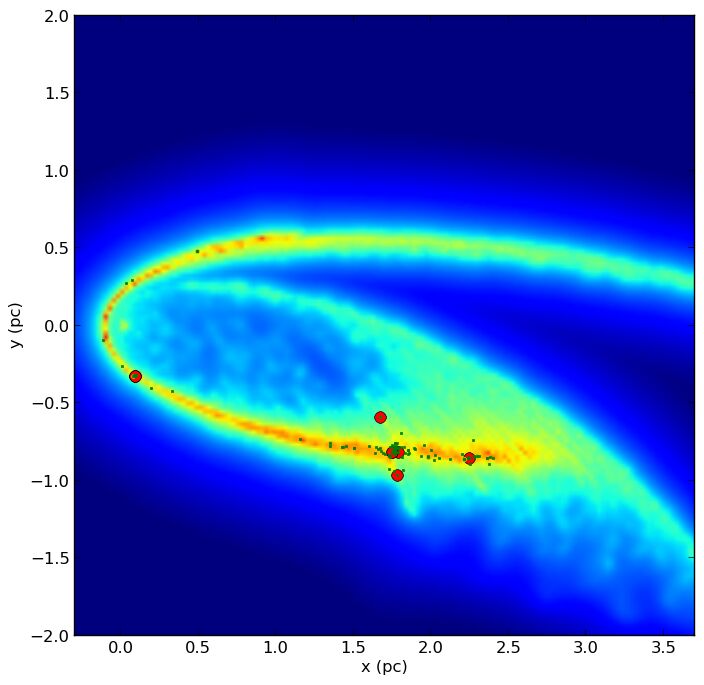}
\includegraphics[width=0.5\textwidth]{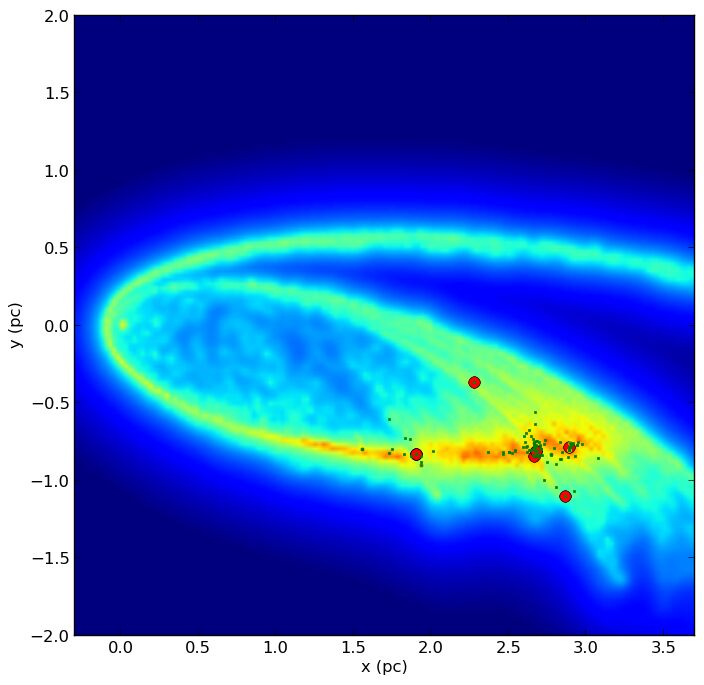}
}
\vspace{-5pt}
\caption{The temporal evolution of column density map for gc10Ke0.944 orbiting model. The temporal snaps and color codes are as in figure~\ref{iso10K}. The clump is rotating anti clockwise. The column densities span logarithmically over $\sim$17.0 to 24.0~$\frac{amu}{cm^{2}}$. The orbit is designed such that the second peri-centre passage occurs at $\sim1.25\times~t_{ff}$. Shown above is the beginning of second orbit of the clump from $\sim3.5$~pc away from the SMBH on the x-axis. After the first peri-centre passage (not shown here) some low-mass protostars have been already formed (caused by the orbital compression), and start accreting more mass toward the second peri-centre passage. A few compact stellar groups (similar to the observed IRS~13N) form around $\sim1.2~t_{ff}$ as discussed in Section 3.3.}
\label{gc10Ke0.944}
\end{figure*}

\begin{figure*}
\centering
\resizebox{0.65\textwidth}{!}{
\vspace{-30pt}
\includegraphics[width=0.5\textwidth]{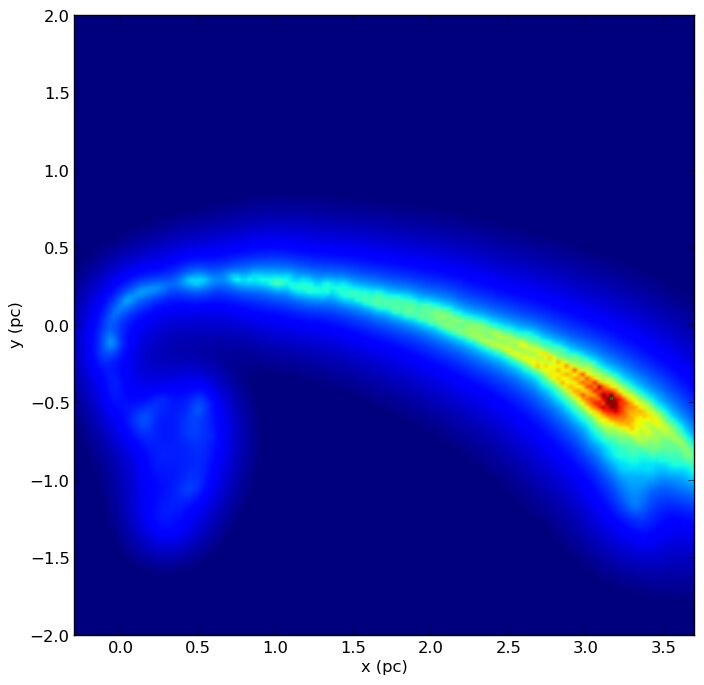}
\includegraphics[width=0.5\textwidth]{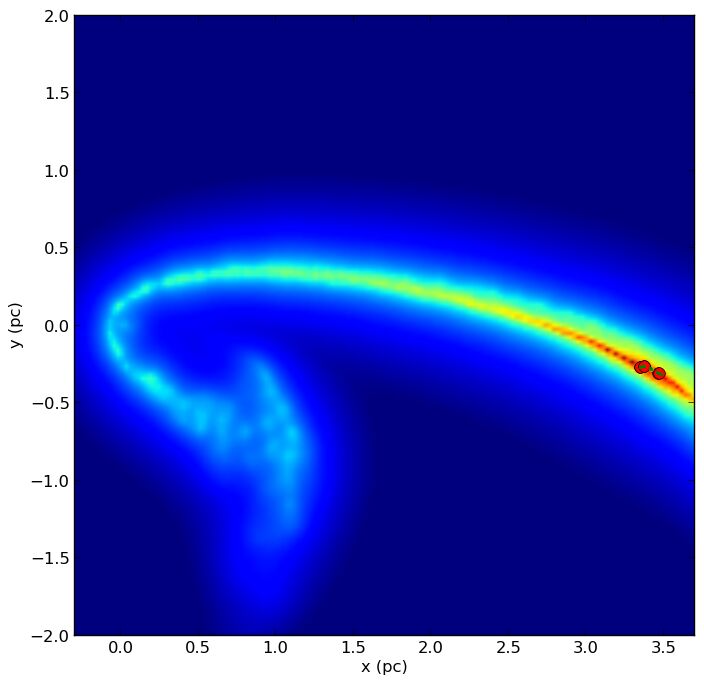}
}
\resizebox{0.65\textwidth}{!}{
\vspace{-3pt}
\includegraphics[width=0.5\textwidth]{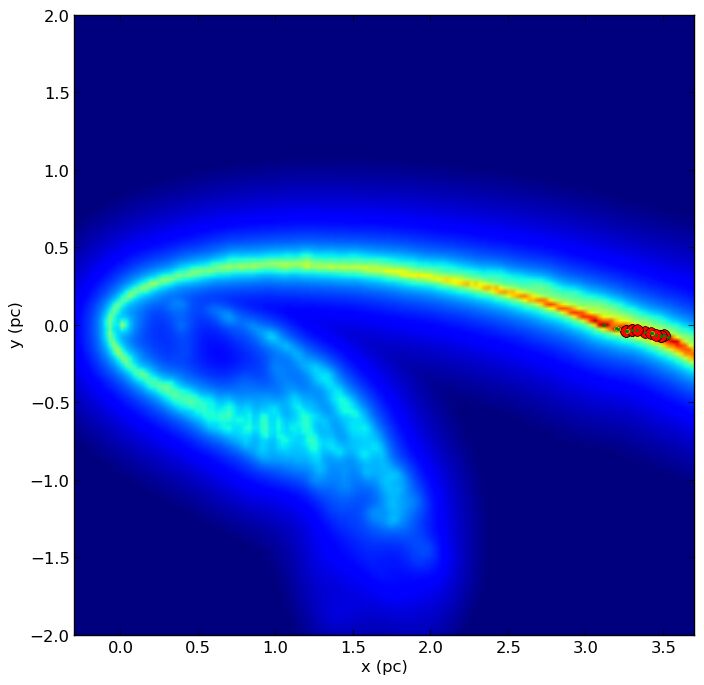}
\includegraphics[width=0.5\textwidth]{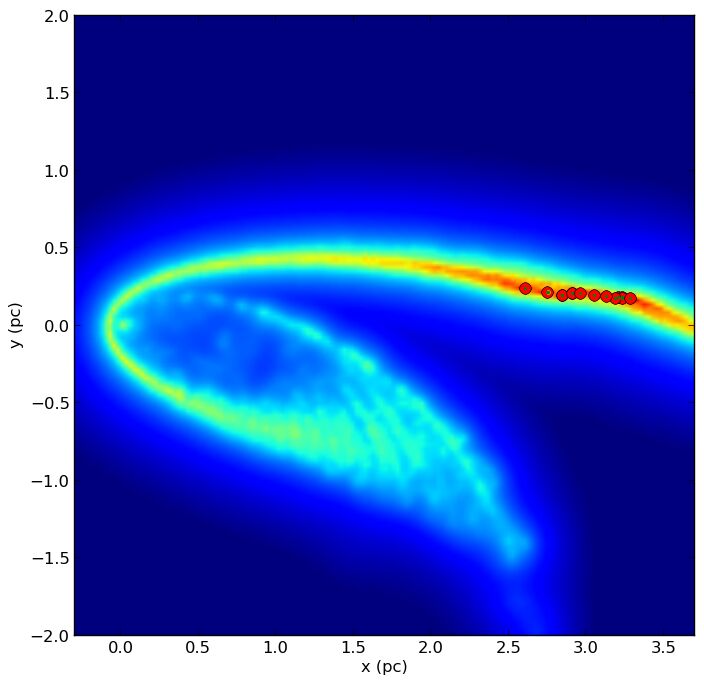}
}
\resizebox{0.65\textwidth}{!}{
\vspace{-3pt}
\includegraphics[width=0.5\textwidth]{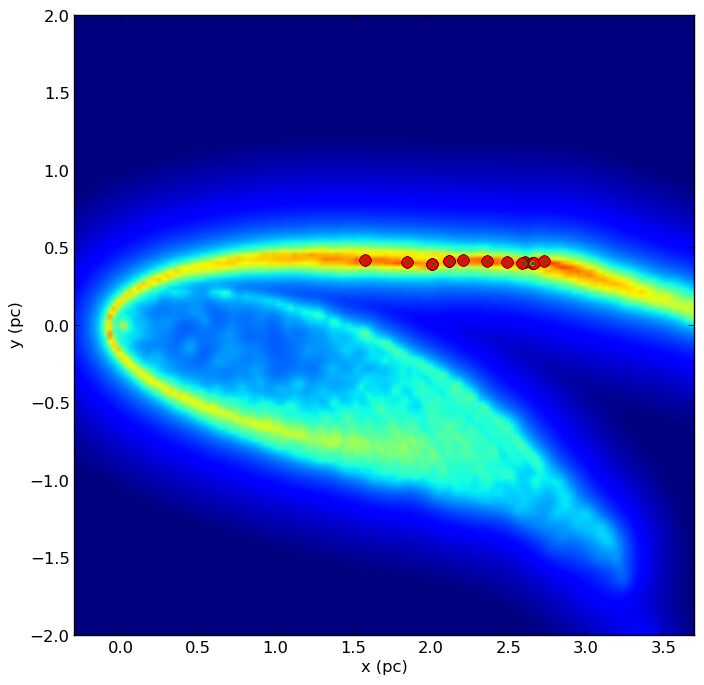}
\includegraphics[width=0.5\textwidth]{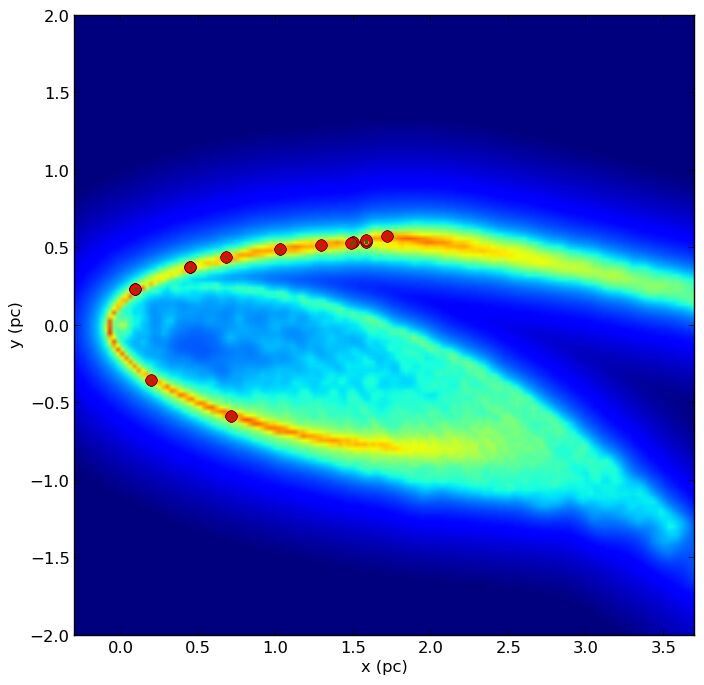}
}
\resizebox{0.65\textwidth}{!}{
\vspace{-3pt}
\includegraphics[width=0.5\textwidth]{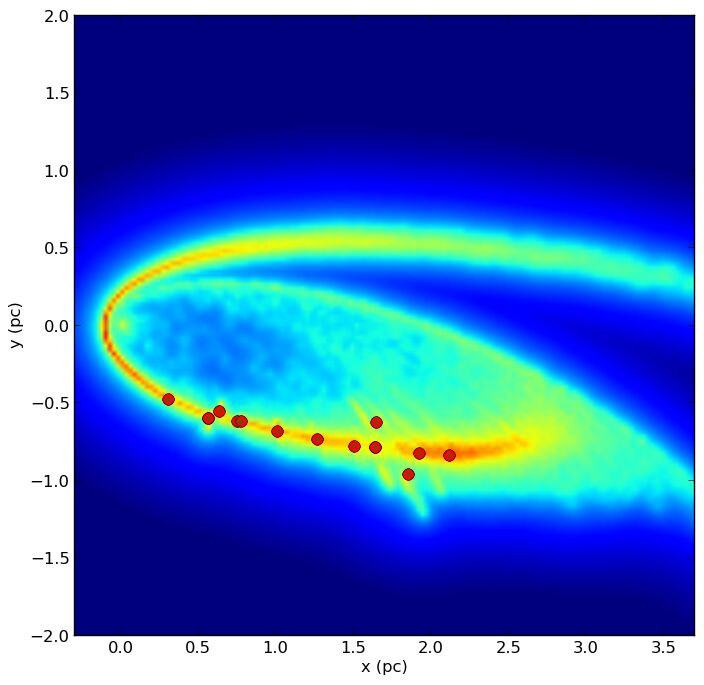}
\includegraphics[width=0.5\textwidth]{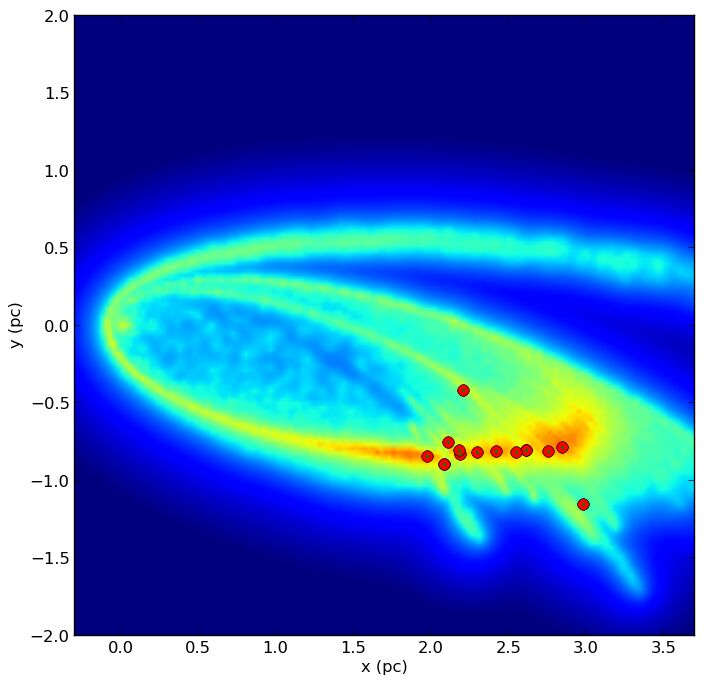}
}
\vspace{-5pt}
\caption{The temporal evolution of column density map for gc50Ke0.944 orbiting model. The temporal snaps and color codes are as in figure~\ref{iso10K}. The column densities span logarithmically over $\sim$17.0 to 24.0~$\frac{amu}{cm^{2}}$. The orbital set up is the same for all the orbiting models. Here, the Jeans mass is higher but, the clump feels the strong tidal field of the SMBH and, as addressed in Section 3, some very massive protostars as well as a couple of compact stellar groups form around $\sim1.2~t_{ff}$.}
\label{gc50Ke0.944}
\end{figure*}

\begin{figure*}
\centering
\resizebox{0.65\textwidth}{!}{
\vspace{-30pt}
\includegraphics[width=0.5\textwidth]{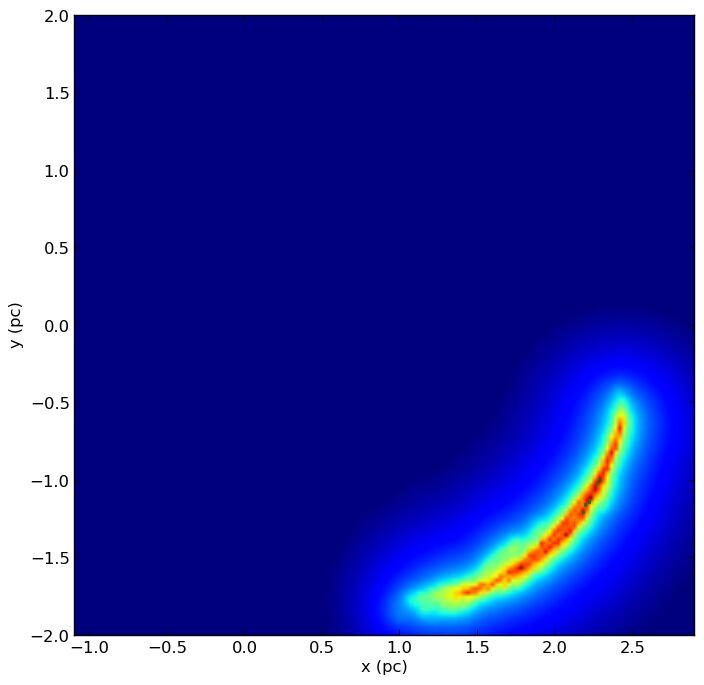}
\includegraphics[width=0.5\textwidth]{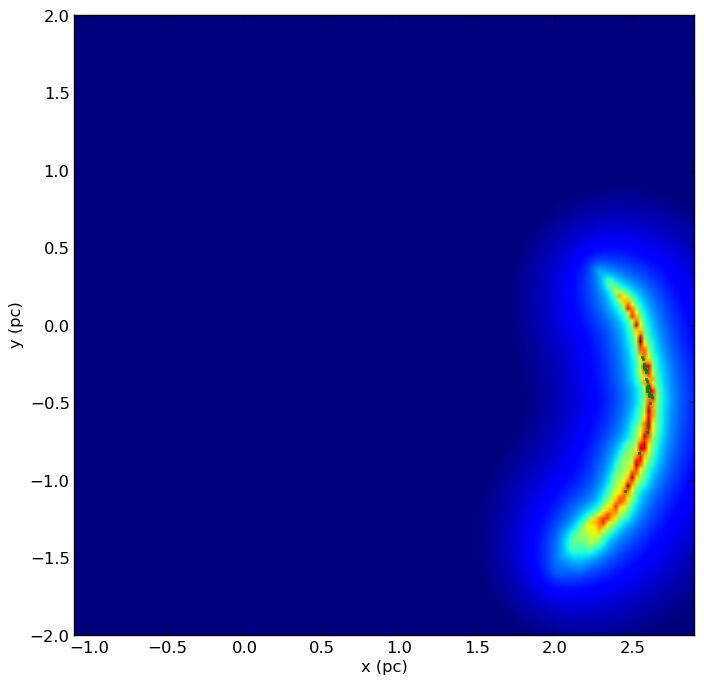}
}
\resizebox{0.65\textwidth}{!}{
\vspace{-3pt}
\includegraphics[width=0.5\textwidth]{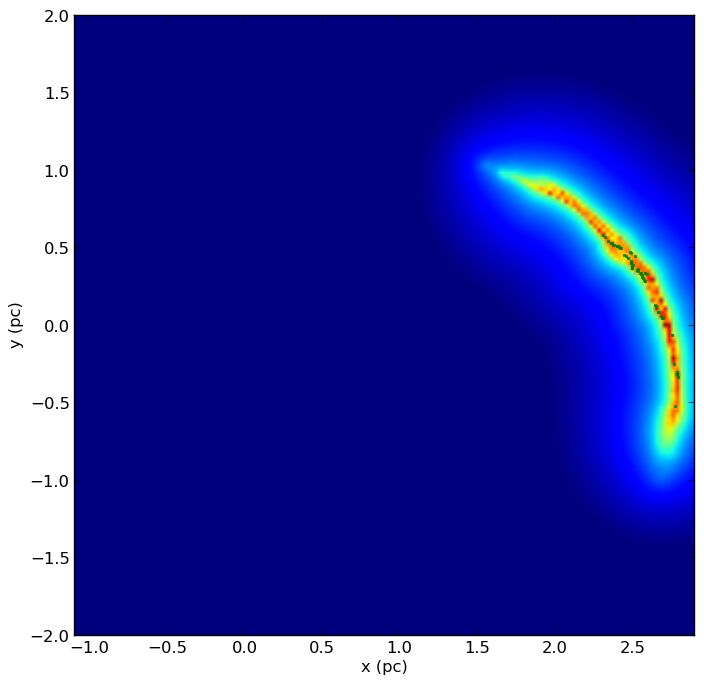}
\includegraphics[width=0.5\textwidth]{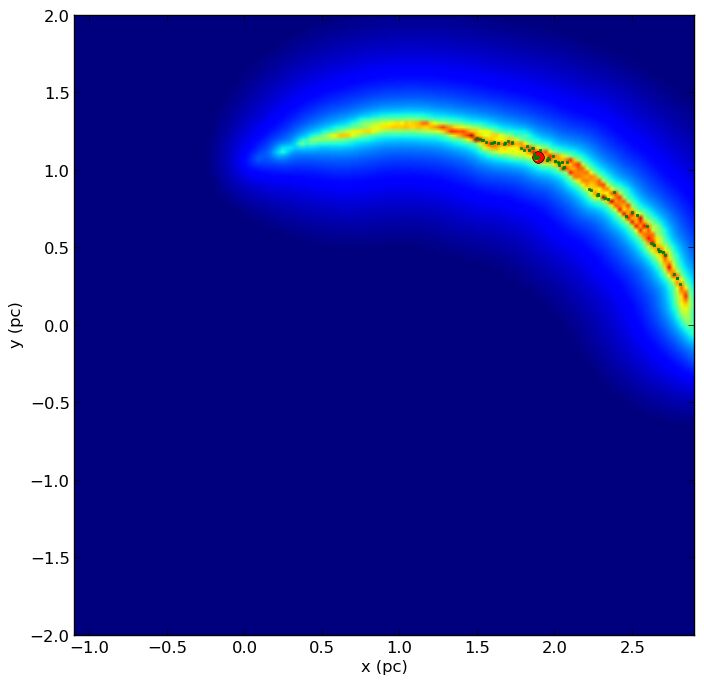}
}
\resizebox{0.65\textwidth}{!}{
\vspace{-3pt}
\includegraphics[width=0.5\textwidth]{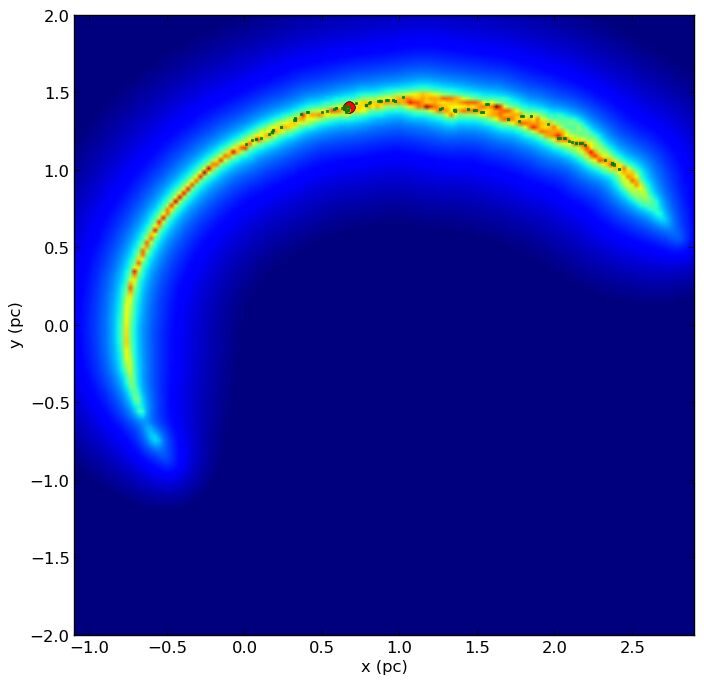}
\includegraphics[width=0.5\textwidth]{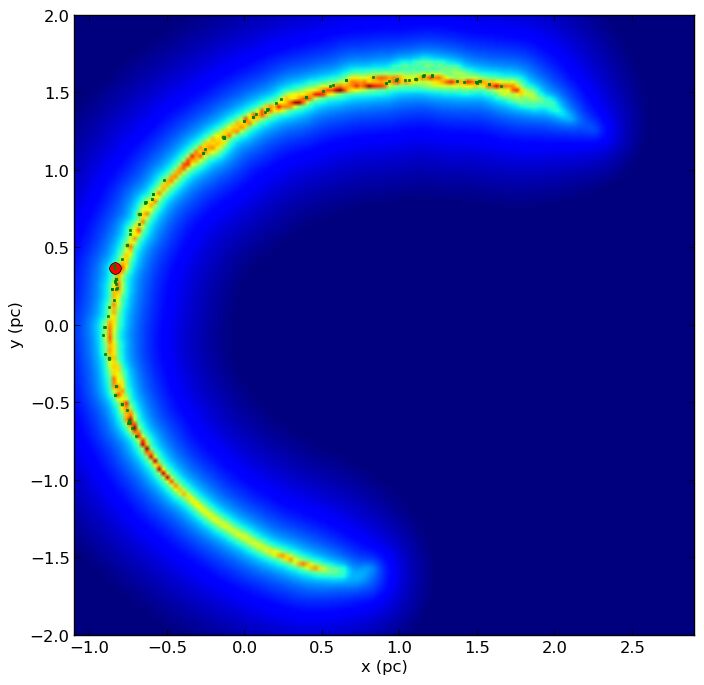}
}
\resizebox{0.65\textwidth}{!}{
\vspace{-3pt}
\includegraphics[width=0.5\textwidth]{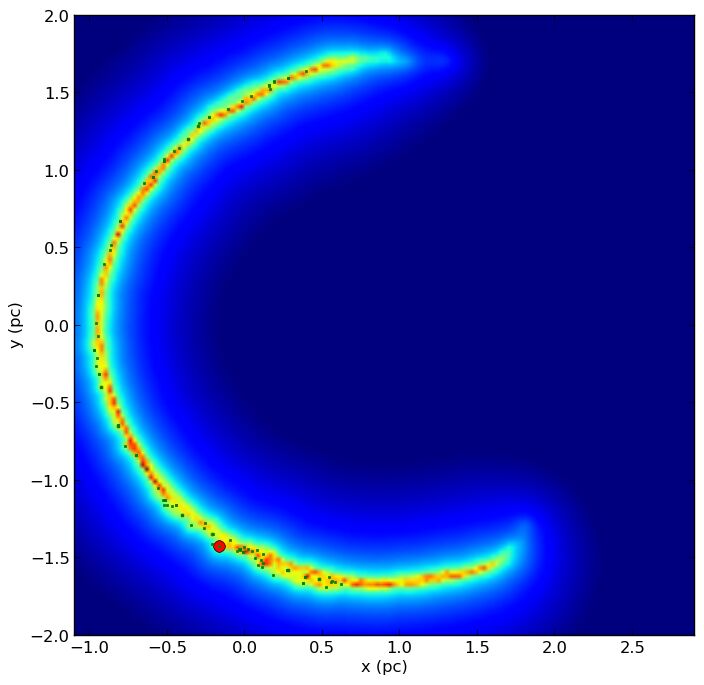}
\includegraphics[width=0.5\textwidth]{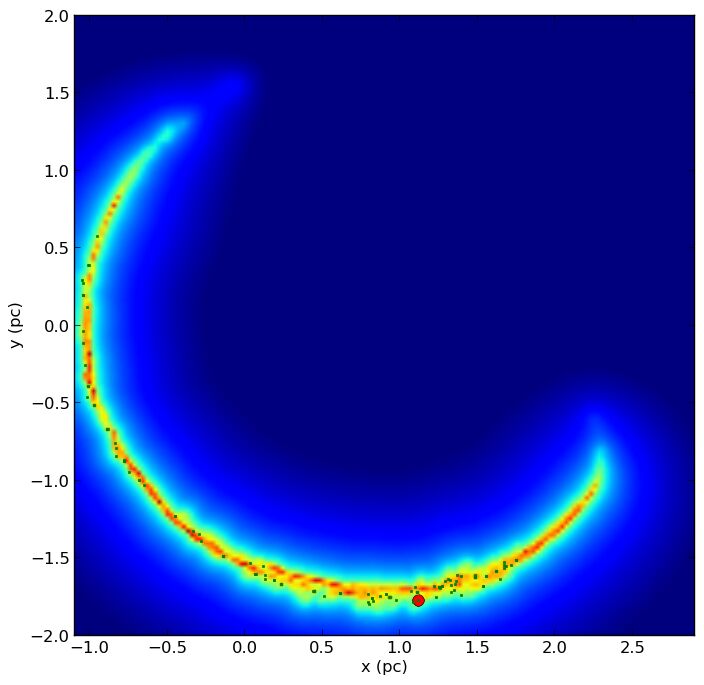}
}
\vspace{-5pt}
\caption{The temporal evolution of column density map for gc10Ke0.5 orbiting model. The temporal snaps and color codes are as in figure~\ref{iso10K}. The column densities span logarithmically over $\sim$16.5 to 24.~$\frac{amu}{cm^{2}}$. The orbital set up is the same as all the orbiting models. There are  several protostars forming in this model as the Jeans mass is as low as in the other 10 K models, also the clump compression along the orbit might help gas densities reach the $\rho_{\mathrm{thresh}}$ limit.}
\label{gc10Ke0.5}
\end{figure*}

\begin{figure*}
\centering
\resizebox{0.65\textwidth}{!}{
\vspace{-30pt}
\includegraphics[width=0.5\textwidth]{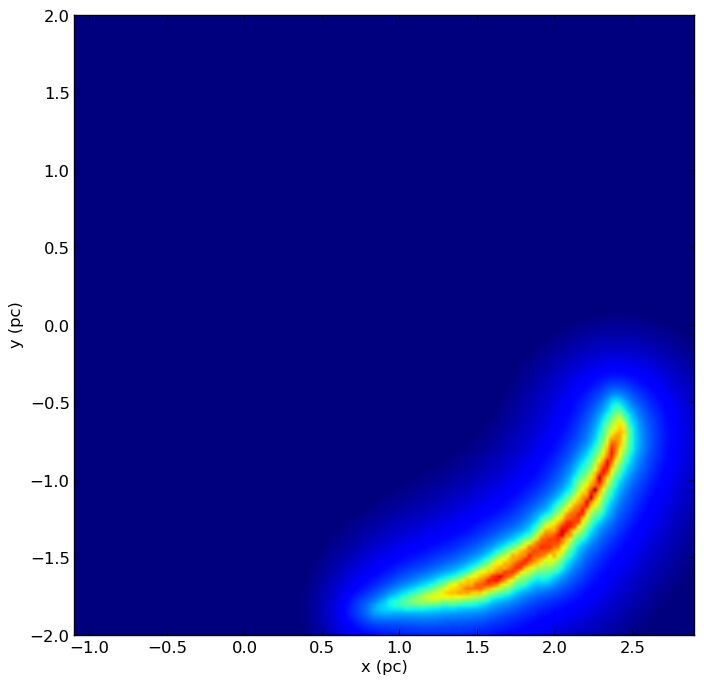}
\includegraphics[width=0.5\textwidth]{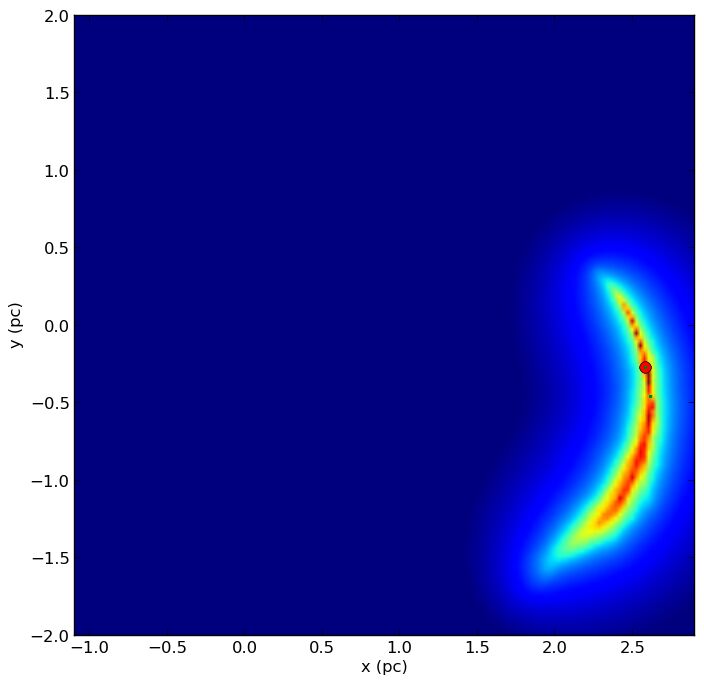}
}
\resizebox{0.65\textwidth}{!}{
\vspace{-3pt}
\includegraphics[width=0.5\textwidth]{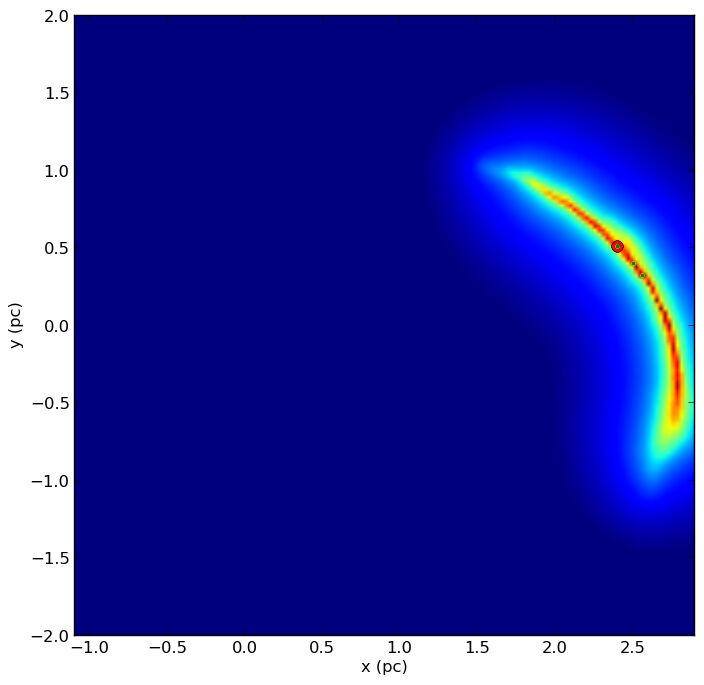}
\includegraphics[width=0.5\textwidth]{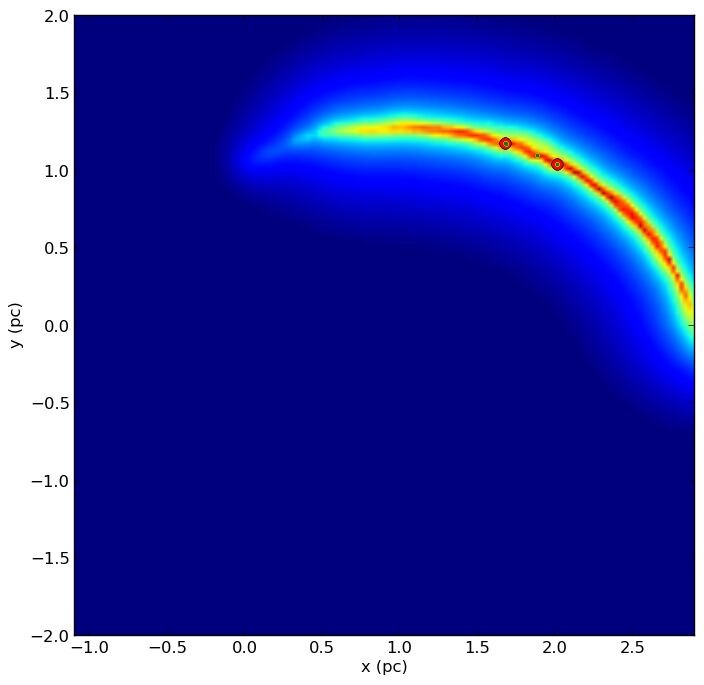}
}
\resizebox{0.65\textwidth}{!}{
\vspace{-3pt}
\includegraphics[width=0.5\textwidth]{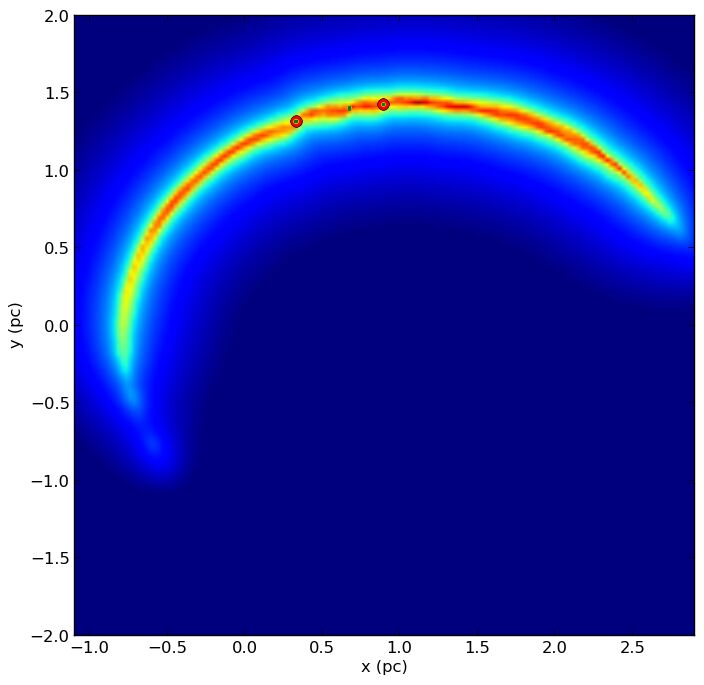}
\includegraphics[width=0.5\textwidth]{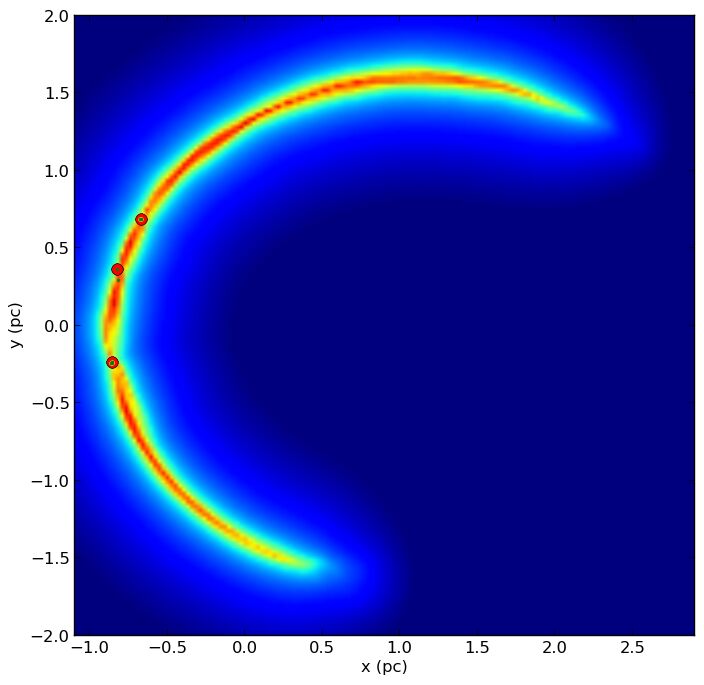}
}
\resizebox{0.65\textwidth}{!}{
\vspace{-3pt}
\includegraphics[width=0.5\textwidth]{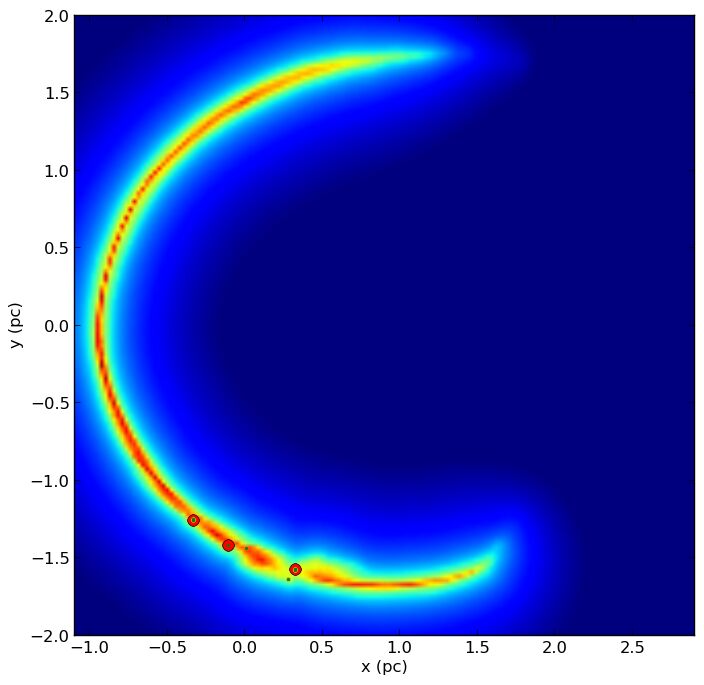}
\includegraphics[width=0.5\textwidth]{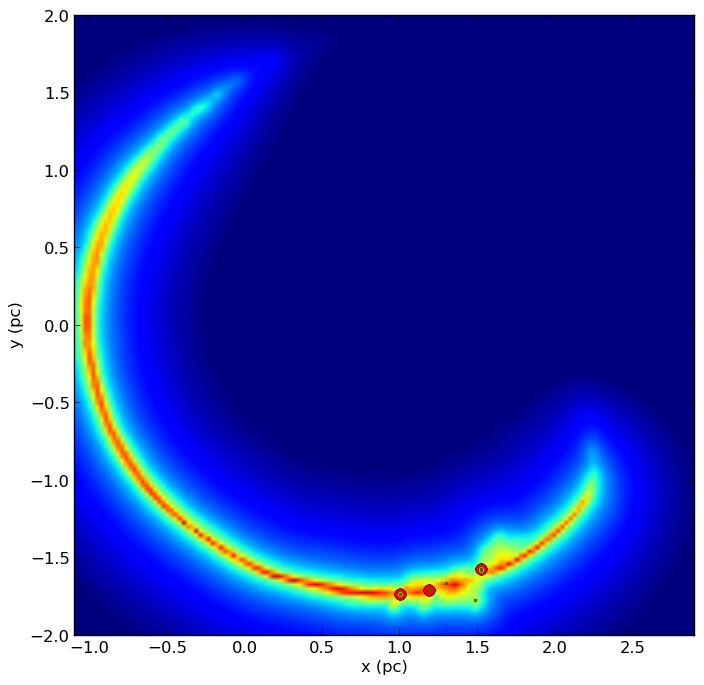}
}
\vspace{-5pt}
\caption{The temporal evolution of column density map for gc50Ke0.5 orbiting model. The temporal snaps and color codes are as in figure~\ref{iso10K}. The column densities span logarithmically over $\sim$16.5 to 24.~$\frac{amu}{cm^{2}}$. The orbital set up is the same as all the orbiting models. There are not as many protostars in this model forming as in its colder counterpart model.}
\label{gc50Ke0.5}
\end{figure*}

The Jeans mass distribution of gas particles for all models as a function of distance to the SMBH (centre of the clump for isolated models) at 1.3~$\times~t_{ff}$ is plotted in Figure~\ref{jeans}.

\subsection{Search for Compact Stellar Systems}

To probe if a structure with similar properties to those observed in IRS~13N forms in our models, we look for a group consisting of a few stars in a region as compact as 0.02~pc in radius and 0.1-0.2 pc away from the SMBH. We report detecting such a group if there are more than two massive stars or if there are two very massive stars ($>2~M_{\odot}$). \citet{Eckart04, 2013A&A...551A..18E} suggest Herbig Ae/Be stars ($\sim$2-8~$M_{\odot}$) as a possible origin of IRS~13N sources. In our current modelling, which does not resolve the opacity limit, it is likely that a very massive star actually consists of a few less massive stars. This makes the comparison between model and observations less precise. Clearly, more sophisticated simulations (including polytropic equation of state and radiative transfer) are needed to investigate this effect better.
We search over the interval of 1.4-1.6 times the orbital period when the clumps are about to pass the peri-centre for the second time. We do not consider low-mass stars ($0.1-0.5~M_{\odot}$) in defining the above group. 

There are in fact more than one such compact small stellar groups in both the highly eccentric models gc10Ke0.944 and gc50Ke0.944. We list the mass and locations of those stars in Table~\ref{table4}.

We only analyze the orbiting models with eccentricity$\sim0.944$ (i.e., $r_{\mathrm{min}}\sim0.1$~pc) to look for IRS~13N-like associations, as in less eccentric models ($r_{\mathrm{min}}\sim0.9$~pc) stars are not close enough to the SMBH as the observed IRS~13N. However, it is interesting to note that there are stars (in 0.5 - 1.2~$M_{\odot}$ mass range) in a compact format in the less eccentric model as well, but not as massive as in the case of the highly eccentric models.

\section{Discussion} 

In this Section we discuss the rate of clumps collision in the CND and the mechanisms that could cause clumps to lose angular momentum and in-spiral toward Sgr~A*. Such mechanisms may also explain the episodic star formation very close to Sgr~A*. We also suggest a possible scenario for the origin of the recently observed G2/DSO infrared object towards the central black hole.

\subsection{Collisions/Encounters of clumps in the CND}

An estimate on the number of CND clumps that can be transferred to highly elliptical orbits towards Sgr~A* may be calculated via estimating the 
collisional rate between the clumps and also by comparing the CND versus the mini-spiral lifetime. We show below that both approaches deliver consistent results.

Following the first approach, observations have shown that the CND is a clumpy environment (e.g. \citealt{1993ApJ...402..173J, 2005ApJ...622..346C, 2013ApJ...770...44L}). There is also strong evidence that shocks and clump-clump collisions are on going within the CND based on the observed $H_{2}$ and HCN emissions (\citealt{2001ApJ...560..749Y, 2013ApJ...767L..32Y}).

Here, we estimate the rate of clumps collision, $\Gamma$, based on the CND structure and properties listed in Table 2 of \citet{2005ApJ...622..346C} using the following relation:
\begin{equation}
\Gamma ~ \sim n \cdot \sigma_v \cdot A_{\mathrm{cross}} \;, 
\end{equation}
here $n$ is the clump's number density, $\sigma_v$ is (one sigma) velocity dispersion of clumps with respect to each other,
and $A_{\mathrm{cross}}$ is the (area) cross section of a typical clump within the CND. 

We assume at least 100 clumps each with 0.25~pc diameter are within the CND central 2~pc with average thickness of 0.5~pc. 
\citet{2005ApJ...622..346C} mention that they observe numerous clumps, but only consider 26 clumps that are isolated in both position and  velocity space in their data. We also estimate, using the above mentioned data, that $\sigma_v$ is about 20 km/s (consistent with \citealt{2010RvMP...82.3121G}) or 2$\times$10$^{-5}$ pc/yr. 

Thus, clumps collision rate can be approximated as:\\
\begin{equation}
\Gamma \sim (\frac{100}{0.5\times2^2}) \cdot (2\times10^{-5}) \cdot (\frac{0.25}{2})^2 ~ \sim 10^{-5}~\mathrm{yr}^{-1}.
\end{equation}

Furthermore, the CND orbital period can be estimated as $10^{5}$~yr, and its lifetime as $10^{7}$~yr (\citet{2005ApJ...622..346C} and references therein). Based on these timescales and our collision rate estimate we argue that during the CND lifetime about 100 collisions are possible. This shows that the CND is an interacting region, and collisions as well as encounters between clumps are common. 

This number of collisions, however, could be in any direction, but to deviating from the disk bulk rotation it is necessary that a clump loses angular momentum to fall in  toward the SMBH (possibly with several encounters). Thus we divide the above number of collisions by two (two out of four possible directions, along/opposite the disk rotation and inward/outward in the disk, assuming the CND is flat with respect to the clump cross-section), about 50 collisions/encounters that are suitable to dissipate orbital energy and trigger inward motion of the corresponding clumps.
About 30\% of those 50 encounters with proper directions will have velocities of more than one-sigma velocity dispersion of the clumps among each other. This leads to about 15 collisions per CND lifetime that efficiently remove angular momentum from clumps and result in highly elliptical clump orbits towards the central SMBH.

As for the second approach, the CND and mini-spiral lifetime also allows us to estimate the number of CND gas and dust entities being transferred onto elliptical orbits. The fact that there are a few streamers (called mini-spiral) in non-circular orbits toward the SMBH is the observational evidence for likely encounters that can lead to this phenomenon. The ratio of the CND lifetime ($10^6$ - $10^7$ yr) over the observed mini-spiral lifetime ($10^4$ - $10^5$ yr) is about 10 to 100. This corresponds to the probability of clump collisions and inward motions. This is in good agreement with our previous estimate 
of $\sim$15 collisions between clumps during the CND lifetime.

The above scenario is also consistent with the possible transient nature of the CND and indications of star formation in it. 
\citet{2012A&A...542L..21R} analyze CO transitions and conclude that the CND has a transient nature and contains warmer and less dense clumps. 
This study suggests that the CND total mass is a few $10^4 M_{\odot}$, one order of magnitude smaller than the mass reported in \citet{2005ApJ...622..346C}. This would consequently imply that the CND has a shorter lifetime of $\sim$~1 Myr. Recently, \citet{2013arXiv1307.8443L} use SOFIA/FORCAST imaging data and conclude on the unstable structure of the inner ring of the CND (they called circumnuclear ring). Their high spatial resolution data confirm the high degree of symmetry of the CND (as in \citet{1993ApJ...402..173J} results), suggesting that the CND has orbited the Galactic centre several times in its lifetime. In this case the authors report that the present-day clumpy structure of the CND (if transient) is puzzling. \citet{2008ApJ...683L.147Y}, on the other hand, detect methanol masers in some of the clumps and interpret it as a signature of early massive star formation (few $10^4$ yr). 

Whether the different tracers provide consistently reliable information on the stable or transient nature of the CND and 
the early phase of star formation probably needs more investigation. One should also consider other mechanisms such as shocks, 
collisions and UV radiation as well as the role of orbital compression close to the SMBH as each effect might act as 
an external pressure on some clumps that could then cause their densities to be high enough to maintain the star formation conditions at least over a specific time interval during the CND lifetime.

\subsection{Young stars close to Sgr~A*: massive binaries and compact groups}

We have shown that, under some observationally supported assumptions about the gas reservoir surrounding the SMBH, some molecular clumps (within the CND) could experience encounters and collisions, losing angular momentum and falling towards the SMBH on eccentric orbits. Our model clump, on its $e=0.944$ orbit is then a simplified version of a more complex picture in which multiple encounters together with the orbital compression increase the gas densities above a threshold value at which the cloud collapses and forms stars close to the SMBH. This mechanism may also explain the presence of small stellar groups very close to the SMBH. Such systems will not be stable over one or two complete orbits around the SMBH and will dissolve quickly. The formed groups of stars such as those we tabulated in Table 4, could evolve due to the internal dynamics of the stellar group. Although having formed from a bound orbiting clump they may dissolve due to internal stellar dynamics. However, depending on the central density of the system some groups such as IRS~13E or IRS~13N can survive long enough that the stellar members reach a more evolved phase.

Some very massive YSOs ($>$20~$M_{\odot}$) recently observed \citep{2013ApJ...767L..32Y} about 0.4~pc away from Sgr~A* could be also explained as protostars forming in a way similar to our modelling, but perhaps starting with more massive initial clumps.

Furthermore, we report that some massive stars (1-5 $M_{\odot}$ each) formed in our simulations are probably binaries similar to those that were recently detected \citep{2013arXiv1307.7996P} close to Sgr~A*. Note that we have used for our present simulations a relatively small and not very massive clump, it is therefore likely that more massive binaries and stellar groups might form if we follow the evolution of a more massive clump.

\subsection{On the origin of the G2/DSO source}

\citet{2012Natur.481...51G} detect an infrared bright object, they called G2, infalling towards Sgr~A* on a highly eccentric orbit \citep{2013arXiv1306.1374G, 2013arXiv1304.5280P}. \citet{2013A&A...551A..18E} present a NIR K-band identification of the object, and called it Dusty S-cluster Object (DSO). The K-band emission implies that it is due to photospheric emission from an embedded star.
The nature of this interesting source is not completely clear, particularly it is yet unknown if the dusty object could survive on its orbit all the way to the vicinity of the SMBH.

There are several models proposed for the nature of G2/DSO. In some it is suggested that a star is necessary to bring the 
observed small cloud in such a harsh environment close to the SMBH. \citet{2013A&A...551A..18E} speculate that the DSO is potentially a young stellar object. \citet{2012NatCo...3E1049M} explore the possibility of a proto-planetary disc surrounding a low-mass star that scattered out of an outer disk of stars towards the SMBH. \citet{2013ApJ...768..108S} suggest that the (mass-loss) envelope of a young T-Tauri star is possibly the origin of the observed source. Future observations can help distinguishing between different scenarios.

Considering the above observational information, we suggest that some of the stars which form  in our highly eccentric models 
and pass close to the central SMBH could be possible G2/DSO progenitors. Such stars could have been formed at any time while the parent 
clump is orbiting the SMBH. As it could be seen in the column density maps of the highly eccentric models (in Figs. 7 and 8), there are a 
few young stars (marked in red) that pass close to the central black hole around the 2nd peri-centre. Additionally, there 
are also a couple of stars that by possible ejection mechanisms are moving towards the SMBH (last raw in Figs. 7 and 8). 
Thus, supported by the results of our simulations, we propose that if young stars could form or pass close to the 
SMBH similar to our models, some of them might be G2/DSO progenitors.

The similar idea of a disrupting star close to Sgr A* as a possible G2/DSO 
progenitor has been recently proposed, e.g. see figure 1 in \citet{2014ApJ...786L..12G}. Such stars could have formed similar to our proposed scenario i.e., via small clumps originated from the CND. Furthermore, \citet{2013arXiv1312.1715M} recently pointed out that a few other sources similar to G2/DSO have been observed further away from Sgr A*. This could be very interesting in the line of our proposed scenario, as in their Fig 2 (L and K composite image), sources 1 and 3 (similar to G2/DSO) are located north of the IRS 13N group. This suggests that young stars, along with their associated dusty envelopes, could sporadically form and pass in the immediate vicinity of the central black hole. Our simulations support such a scenario.

\section{Summary and Conclusions}

Our main results can be summarized as follows:\\
~\\
1. In our simulations, highly eccentric model clumps, starting with a moderate density ($10^5$ cm$^{-3}$), passing very close to the SMBH could evolve to high enough densities that are higher than the SMBH's tidal densities and are necessary for star formation to occur in our approach. Particularly for the highly eccentric models, orbital compression of the clumps perpendicular to the orbit is the dominant factor while the clump is stretched along the orbit (Figure~\ref{gas_sink}).\\ 
~\\
2. A few massive young stars, some in the form of massive binaries, formed in all the model calculations regardless of the presence of the SMBH. As expected, more massive stars (usually less numerous) form in the case of warmer clumps consistent with the higher Jeans mass criterion (Figure~\ref{imf100msun} top panel).\\
~\\
3. In the case of the two high eccentric models (Table~\ref{table4}) a few compact young stellar groups with similar properties as the observed IRS~13N group form. Thorough modelling is planned to further investigate the formation and properties of such young compact systems. Specifically, including a more realistic equation of state for the molecular clump as well as resolving the opacity limit are necessary for better comparison of the mass and number of protostars with the observations.\\
~\\
4. In our present modelling, the mass distributions (IMF) of newly formed stars, in almost all models, are consistent with the \citet{2001MNRAS.322..231K} IMF (Figure~\ref{imf100msun} bottom panel).\\
~\\
5. The scenario described in this paper provides a model that can explain the ongoing formation of stars and small stellar 
clusters in the Galactic centre region, in addition to the episodic massive star formation as also explained by \citet{2005A&A...437..437N}.\\
~\\
6. We show that clumps could encounter with each other in the CND environment and it is plausible that some clumps, as the one modelled in our simulations, eventually lose angular momentum and fall in towards Sgr~A*. The observed streamers in non-circular orbits about the central black hole  could be explained by this argument. Collisions of molecular clumps also suggest that external pressure on clumps could help trigger massive star formation within the clumps in situ close to the SMBH (Section 4.1).\\
~\\
7. Finally, we propose a scenario for the origin of the newly discovered G2/DSO source falling towards the peri-centre point 
of a highly elliptical orbit around the SMBH. Based on our simulations, we suggest that a few young (embedded) stars could form close to the SMBH within a small star forming molecular clump, such as those highly eccentric clumps modelled in our present work (Section 4.2). Such stars at some point could pass very close to the SMBH on eccentric orbits (e.g., around the second peri-centre in our models). It is then possible that a dusty gaseous part of such a star could be separated (at least temporarily) from its stellar progenitor.\\

\section*{Acknowledgments}

We acknowledge the anonymous referee for the constructive comments that helped clarifying the manuscript.
We thank the regional computing center of the university of cologne (RRZK) for providing computing time on the DFG-funded High Performance Computing (HPC) system CHEOPS. This work was also supported by the Netherlands Research Council NWO (grants \#643.000.803 [IsFast], \#639.073.803 [VICI], \#614.061.608 [AMUSE] and \#612.071.305 [LGM]), by the Netherlands Research School for Astronomy (NOVA). 
B.J. is grateful to Volker Winkelmann kind support at the HPC center, and to Arjen Elteren and Nathan de Vries friendly help on the AMUSE installation and its functionality issues. B.J. thanks Fatemeh Motalebi for helping on Figure~\ref{scheme} visualization to be as informative as possible. We thank Mattew Horrobin for English revision. 
This work was supported in part by the Deutsche Forschungsgemeinschaft (DFG) via the Cologne Bonn Graduate School (BCGS), 
and via grant SFB 956, as well as by the Max Planck Society and the University of Cologne through 
the International Max Planck Research School (IMPRS) for Astronomy and Astrophysics.
We had fruitful discussions with members of the European Union funded COST Action MP0905: Black Holes in a violent Universe and the
COST Action MP1104: Polarization as a tool to study the Solar System and beyond.
We received funding from the European Union Seventh Framework Programme (FP7/2007-2013) under grant agreement No.312789.

\bibliographystyle{mn2e}
\bibliography{BJalali_IRS13N}

\section{Appendix: Orbital Compression}

\begin{figure}
\includegraphics[width=0.9\columnwidth]{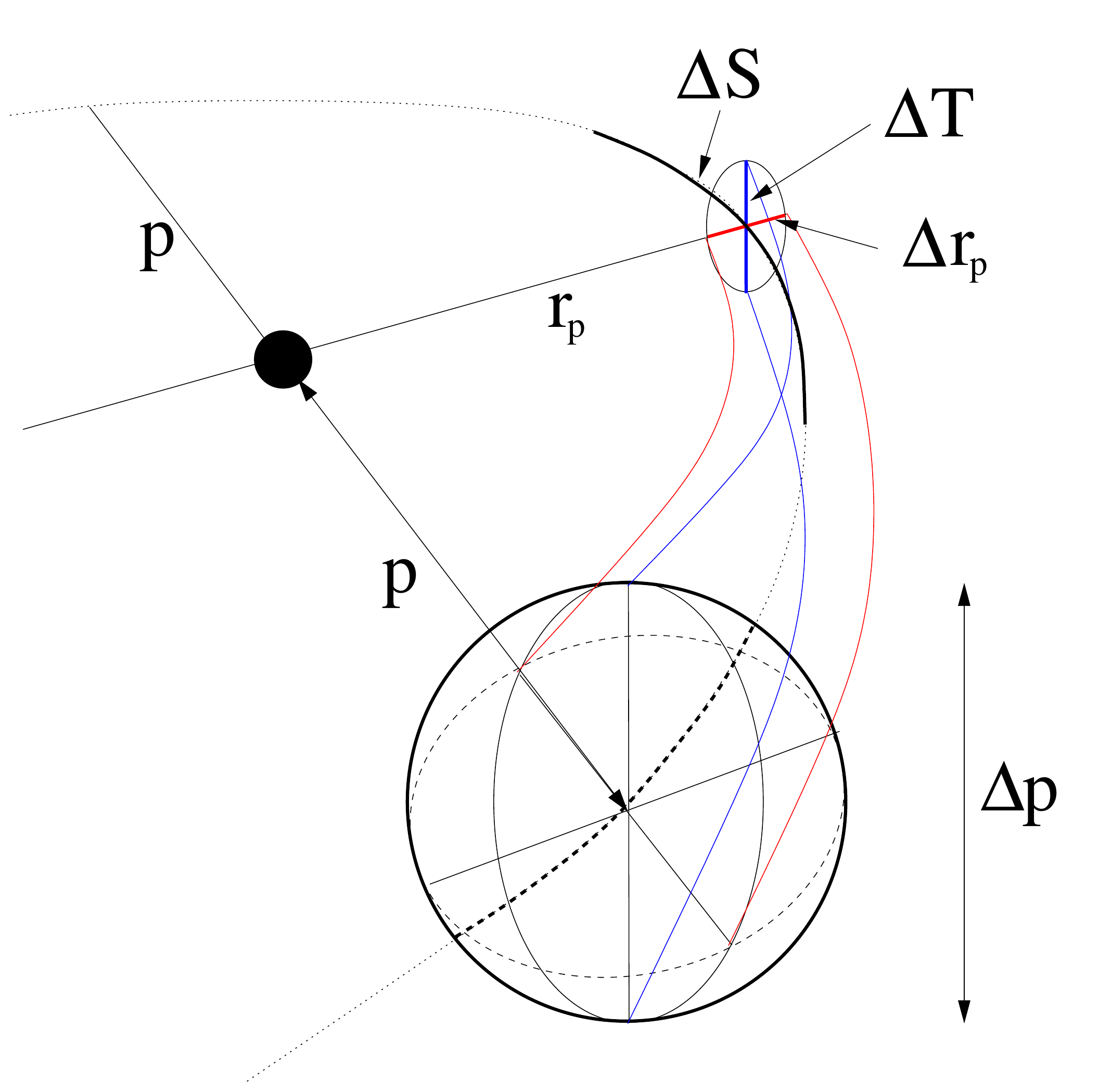} 
\caption{Schematic structure of a source going under orbital compression close to the peri-centre passage.
The filled black point is the central black hole, p and $r_P$ are the semi-latus rectum and peri-centre distances. Thin blue and red lines as well as the
 thin black line at the peri-centre are to guide the reader on stretching the cross-sections of the source along the orbit.}
\label{schematic}
\end{figure}

In order to discuss the orbital compression of a source on an elliptical orbit around a 
black hole we consider a source with diameter $\Delta p$ at a position along the orbit 
that touches the semi-latus rectum, i.e. the source is at the location of the black hole 
if one projects its location perpendicular onto the semi-major axis.
From there it continues to descend towards the peri-centre. At the semi-latus rectum 
location we approximate the source volume via
\begin{equation}
V_1 \cong (\Delta p)^3~~.
\end{equation}
In Fig.~\ref{schematic} we show a source with diameter $\Delta p$ at its semi-latus rectum position. The three perpendicular
cross-sections of the source stretch along the orbit and are compressed along the red and blue thin lines
to the thick black, red and blue cross-sections of the orbitally compressed source at the position of the
peri-centre $r_P$.
Following Kepler's second law we find 
\begin{equation}
p~\Delta p \cong r_P~\Delta S~~.
\end{equation}
This implies that the object is stretched along the orbit and the size $\Delta p$ turns into
\begin{equation}
\Delta S \cong \frac{p~\Delta p}{r_P}~~.
\end{equation}
The peri-centre distance $r_P$ is related to the ellipticity $e$ and the semi-latus rectum distance via
\begin{equation}
r_P = \frac{p}{1+e}~~.
\end{equation}
Therefore, in the orbital plane a variation of $p$ that corresponds to the diameter 
of the object $\Delta p$ translates into
\begin{equation}
\Delta r_P = \frac{\Delta p}{1+e}~~.
\end{equation}
All particles of the object that are located above and below the principle orbital plane are focused towards
an intersection line in this plane centred at the peri-centre location. Therefore we can safely assume $\Delta r_P$
to be an upper limit of the source diameter $\Delta T$ perpendicular to the orbital plane
at the peri-centre location:
\begin{equation}
\Delta T \le \Delta r_P ~~. 
\end{equation}
Hence, the volume of the source at peri-centre can be approximated via
\begin{equation}
V_2  \le \Delta S~\Delta T~\Delta r_P 
= \frac{p~\Delta p}{r_P} (\Delta r_P)^2
= \frac{p~\Delta p (1+e) (\Delta p)^2}{p (1+e)^2}
= \frac{(\Delta p)^3}{1+e}~~.
\end{equation}
This has to be compared to the volume $V_1$ and for $e \rightarrow 1$ we find 
\begin{equation}
\frac{V_2}{V_1} \le \frac{(\Delta p)^3}{(\Delta p)^3} \frac{1}{1+e} = \frac{1}{1+e} < \frac{1}{2}~~.
\end{equation}
The value of $1/2$ is an upper limit. The value can be much smaller depending on how much
the gas volume can be compressed perpendicular to the orbital plane towards the peri-centre.
Part of this compression will be balanced by the internal pressure of the cloud or
turbulences that will be generated close to peri-centre due to shearing motions in azimuthal direction.
Particularly for small sources or small sections of a source with $\Delta p << r_P$ the shear effects 
and possible velocity gradients on the source along the orbit result only in differential effects on
the corresponding volume compression.

\bsp
\label{lastpage}

\end{document}